\documentclass[a4paper,11pt]{article}
\pdfoutput=1 \usepackage{amssymb}
\usepackage{amsmath}
\usepackage{braket}
\usepackage{mathtools}
\usepackage[usenames,dvipsnames,svgnames,table]{xcolor}
\usepackage[utf8]{inputenc}
\usepackage{color}
\usepackage{cite}
\usepackage{subcaption}
\usepackage{lmodern}
\usepackage{footnote}
\usepackage[normalem]{ulem}
\usepackage{glossaries-extra}
\setabbreviationstyle[acronym]{long-short}
\glssetcategoryattribute{acronym}{nohyperfirst}{true}

\usepackage{slashed}
\usepackage[pdftex]{graphicx}
\usepackage{multirow}
\usepackage{jheppub}
\usepackage{here}
\usepackage{hyperref}
\usepackage{verbatim}
\usepackage[justification=centering,singlelinecheck=false]{caption}
\usepackage{cleveref}
\usepackage{listings}
\usepackage{fancyvrb}
\usepackage{dsfont}
\usepackage{nicefrac,xfrac}

\Crefname{equation}{eq.}{eqs.}
\Crefname{section}{section}{sections}
\Crefname{figure}{figure}{figures}
\Crefname{appendix}{appendix}{appendices}

\newcommand{\cB}{\mathcal{B}}

\newcommand{\cD}[0]{\mathcal D}

\newcommand{\cK}[0]{\mathcal K}

\newcommand{\cM}[0]{\mathcal M}
\newcommand{\cO}[0]{\mathcal O}

\newcommand{\cS}[0]{\mathcal S}
\newcommand{\cT}[0]{\mathcal T}
\newcommand{\cY}[0]{\mathcal Y}

\newcommand{\wt}[0]{\widetilde}
\newcommand{\wh}[0]{\widehat}

\newcommand{\df}[0]{\mathrm{df}}

\newcommand{\Kdf}[0]{{\cK_{\df,3}}}

\newcommand{\PV}[0]{{\mathrm{PV}}}

\newcommand{\K}[0]{\mathcal K}

\newcommand{\bm}[0]{\boldsymbol}

\newcommand{\bcX}{\boldsymbol{\mathcal X}}
\newcommand{\bcY}[0]{\boldsymbol{\mathcal Y}}
\newcommand{\bcK}[0]{{\overline{\mathcal K}}}

\newcommand{\XR}[3]{\boldsymbol{\mathcal X}_{[kab]}^{(#1#2#3)}}
\newcommand{\YL}[3]{\boldsymbol{\mathcal Y}^{[kab] \dagger}_{(#1#2#3)}}
\newcommand{\YR}[3]{\boldsymbol{\mathcal Y}^{[kab]}_{(#1#2#3)}}
\newcommand{\CL}[0]{\boldsymbol{\mathcal C}^\dagger}
\newcommand{\CR}[0]{\boldsymbol{\mathcal C}}

\newcommand{\KSS}[0]{Kim:2005gf}

\newcommand{\HSQCa}[0]{Hansen:2014eka}
\newcommand{\HSQCb}[0]{Hansen:2015zga}

\newcommand{\BHSQC}[0]{Briceno:2017tce}

\newcommand{\BHSK}[0]{Briceno:2018aml}
\newcommand{\dwave}[0]{Blanton:2019igq}

\newcommand{\largera}[0]{Romero-Lopez:2019qrt}

\newcommand{\isospin}[0]{Hansen:2020zhy}

\newcommand{\implement}[0]{Blanton:2021eyf}
\newcommand{\threeN}[0]{Draper:2023xvu}
\newcommand{\tetraquark}[0]{Hansen:2024ffk}
\newcommand{\BSQC}[0]{Blanton:2020gha}
\newcommand{\BSequiv}[0]{Blanton:2020jnm}
\newcommand{\BSnondegen}[0]{Blanton:2020gmf}
\newcommand{\BStwoplusone}[0]{Blanton:2021mih}

\newcommand{\Akakia}[0]{Hammer:2017uqm}
\newcommand{\Akakib}[0]{Hammer:2017kms}

\newcommand{\MD}[0]{Mai:2017bge}

\newcommand{\kdfnloall}[0]{Baeza-Ballesteros:2024mii}

\newcommand{\Kplus}{K^{+}}
\newcommand{\Kbar}{\overline{K}^{0}}
\newcommand{\Kzero}{K^{0}}
\newcommand{\Kminus}{K^{-}}
\newcommand{\piplus}{\pi^{+}}
\newcommand{\pizero}{\pi^{0}}
\newcommand{\piminus}{\pi^{-}}

\newcommand{\ThreeBodyFormalism}[0]{%
Briceno:2012rv,
Polejaeva:2012ut,
Hansen:2014eka,
Hansen:2015zga,
Briceno:2017tce,
Hammer:2017uqm,
Konig:2017krd,
Hammer:2017kms,
Mai:2017bge,
Briceno:2018aml,
Blanton:2019igq,
Pang:2019dfe,
Jackura:2019bmu,
Romero-Lopez:2019qrt,
Hansen:2020zhy,
Blanton:2020gha,
Blanton:2020jnm,
Pang:2020pkl,
Romero-Lopez:2020rdq,
Blanton:2020gmf,
Muller:2020vtt,
Blanton:2021mih,
Muller:2021uur,
Blanton:2021eyf,
Jackura:2022gib}

\newacronym{CMF}{CMF}{center-of-momentum frame}

\DeclareFixedFont{\ttb}{T1}{txtt}{bx}{n}{9}
\DeclareFixedFont{\ttm}{T1}{txtt}{m}{n}{9}

\definecolor{deepblue}{rgb}{0,0,0.5}
\definecolor{deepred}{rgb}{0.6,0,0}
\definecolor{deepgreen}{rgb}{0,0.5,0}
\definecolor{jlab_red}{RGB}{192,39,45}
\definecolor{jlab_orange}{RGB}{249,102,0}
\definecolor{jlab_blue}{RGB}{47,122,121}
\definecolor{jlab_green}{RGB}{65,125,10}
\definecolor{jlab_blue}{RGB}{47,122,121}

\title{
Three-particle formalism for multiple channels: the $\eta \pi \pi + K \overline K \pi$ system in isosymmetric QCD
}

\author{Zachary T. Draper}
\affiliation{Physics Department, University of Washington, Seattle, WA 98195-1560, USA}
\emailAdd{ztd@uw.edu}

\author{and Stephen R. Sharpe}
\emailAdd{srsharpe@uw.edu}

\abstract{
We generalize previous three-particle finite-volume formalisms 
to allow for multiple three-particle channels.
For definiteness, we focus on the two-channel $\eta \pi \pi$ and $K \overline K \pi$ system in isosymmetric QCD,
considering the positive $G$ parity sector of the latter channel,
and neglecting the coupling to modes with four or more particles.
The formalism we obtain is thus appropriate to study
the $b_1(1235)$  and $\eta(1295)$ resonances.
The derivation is made in the generic relativistic field theory approach
using the time-ordered perturbation theory method.
We study how the resulting quantization condition reduces to that for a single three-particle channel when
one drops below the upper ($K\overline K \pi$) threshold.
We also present parametrizations of the three-particle K matrices that enter into the formalism.
}

\allowdisplaybreaks

\begin{document}

\maketitle
\flushbottom
\clearpage

\section{Introduction}

A multitude of exotic hadrons has been unearthed in the last two decades~\cite{Ali:2017jda,Olsen:2017bmm,Karliner:2017qhf,Guo:2017jvc,Liu:2019zoy,Brambilla:2019esw,Gershon:2022xnn,Lebed:2022vfu}, 
sparking interest in providing first-principles predictions of their properties through quantum chromodynamics (QCD). Despite the theoretical suitability of lattice QCD for such endeavors, its practical application encounters limitations due to the extensive array of open decay modes associated with many of these exotic resonances. 

In this work we address one particular limitation of the formalism that has been derived to 
date~\cite{\ThreeBodyFormalism},
namely the restriction to a single three-particle channel~\cite{\HSQCa,\HSQCb,\Akakia,\Akakib,\MD}, 
or to several degenerate such channels~\cite{\isospin,\tetraquark}.
In particular, we consider a system with two distinct three-particle channels, although the generalization
to multiple such channels is straightforward.
This is a step on the way to a completely general formalism that includes channels with two, three and
more particles.\footnote{%
We note that the formalism that includes channels with both two and three particles has been 
derived in ref.~\cite{\BHSQC}, but this is restricted to the case in which all the particles are spinless and identical.
Another approach to the ``$2+3$'' case has recently been proposed for the $D^* D+ D D \pi$ system
in ref.~\cite{\tetraquark}.}

To have a concrete system to discuss, we consider the $b_{1}(1235)$ and $\eta(1295)$ resonances
in isosymmetric QCD.
These have quantum numbers $I^{G}(J^{PC}) = 1^{+}(1^{+-})$ and $0^{+}(0^{-+})$,
and in particular have positive $G$ parity and unnatural $J^P$.
These resonances decay into two three-particle channels,
namely $K \overline K \pi$ and $\pi\pi \eta$, as well as into $4\pi$ modes.
They cannot decay to two pions (a channel having natural $J^P$) or to three pions (due to $G$ parity).
Thus, if we neglect the four pion modes, 
and also the other kinematically allowed channels that have not yet been observed in the
decays of these resonances ($6\pi$, $8\pi$, $4\pi+\eta$),
this is a system that has our desired property of two different three-particle channels.\footnote{%
The $b_1$ resonance has been studied using lattice QCD for heavier-than-physical quark masses such
that $M_\pi\approx 390\;$MeV, in which case it decays primarily to the two-particle channel $\pi\omega$,
and thus can be analyzed using the two-particle quantization condition~\cite{Woss:2019hse}.
}

In this work we make the desired generalization using the generic relativistic field-theoretic (RFT) approach to the derivation of three-particle formalism~\cite{\HSQCa,\HSQCb}.
In particular, we use the method of derivation based on time-ordered perturbation theory (TOPT)
introduced in ref.~\cite{\BSQC}.
The formalism we obtain is a synthesis of methodologies developed previously
for three distinct particles~\cite{\BSnondegen},
for $2+1$ systems (comprising two identical and one distinct particle)~\cite{\BStwoplusone}, 
for three pions of arbitrary isospin~\cite{\isospin}, 
and for the doubly-charmed tetraquark~\cite{\tetraquark}. 
We avoid repeating steps in the derivation that are presented in these works
to the extent possible while retaining readability.

This paper is organized as follows.
In the following section, \Cref{sec:overview}, we introduce the two-channel system that we study,
and point out the pertinent general features.
The core of this paper is \Cref{sec:topt_derivation}, in which we derive the multichannel
three-particle quantization condition, following what are now fairly standard steps in the TOPT approach.
The only new feature is the projection onto the sector of positive $G$ parity,
discussed in \Cref{sec:G}. 
Since the $\pi\pi\eta$ and $K\overline K \pi$ channels have substantially different thresholds, one can ask
how the two-channel formalism reduces to that for a single channel as one drops below the upper
($K\overline K\pi$) threshold. We address this in \Cref{sec:reduction}.
Practical applications require the use of parametrizations of the three-particle K matrix, and
we describe in \Cref{sec:Kdfparam} how symmetries constrain the allowed forms.
We conclude in \Cref{sec:conc}. 
Various technical details are collected into seven appendices.

\section{Overview}
\label{sec:overview}

The total isospin of both $K \overline K \pi$ and $\pi\pi\eta$ channels can be $I=0$, $1$, or $2$.
Although resonances are present only for the first two of these values, we derive the formalism
for all three choices.
All three can be accessed by studying 
the sector with quantum numbers $I_{3}=0$, $U=0, D=0, S=0$.\footnote{%
We have also checked the final results by considering the channels with
$I_3=1$ and $2$, which contain, respectively, total isospins $I=1,2$ and $I=2$.}
There are six different flavor channels with these quantum numbers,
\begin{equation}
\begin{split}
\bigg\{ \Kplus(\bm k_1) \Kbar(\bm k_2) \piminus (\bm k_3), 
\ \Kplus(\bm k_1) \Kminus(\bm k_2) \pizero (\bm k_3), &
\ \Kzero(\bm k_1) \Kbar(\bm k_2) \pizero (\bm k_3), \\
\ \Kzero(\bm k_1) \Kminus(\bm k_2) \piplus (\bm k_3),
\ \piplus(\bm k_1) \piminus(\bm k_2) \eta(\bm k_3), &
\ \pizero(\bm k_1) \pizero(\bm k_2) \eta(\bm k_3)\bigg\}\,,
\end{split}
\label{eq:flavorchannels0}
\end{equation}
where we include momentum labels for later use.
These decompose into isospin as the $I_3=0$ components of
the following seven channels,
\begin{align}
\begin{split}
I=2:& \quad a)\ [[ K \bar K]_1 \pi ]_2,\  b)\ [[\pi\pi]_2 \eta ]_2 \,;
\\
I=1:& \quad a)\ [[K \bar K]_1 \pi]_1,\ b)\ [[K \bar K]_0 \pi]_1,\  c)\ [ [\pi\pi]_1 \eta ]_1\,;
\\
I=0:& \quad a)\ [[K \bar K]_1 \pi]_0,\ b)\ [[\pi\pi]_0 \eta ]_0\,;
\end{split}
\label{eq:isospindecomp}
\end{align}
where the subscripts indicate isospin, and the momentum labels are implicitly
ordered as in \Cref{eq:flavorchannels0}: $\bm k_1$, $\bm k_2$, and $\bm k_3$.
The increase from six to seven dimensions arises because, when decomposing under isospin,
the symmetric and antisymmetric parts of $\pi^+\pi^-$ are treated separately.
The relation between the isospin and flavor bases, which will be needed
in the subsequent discussion, is given in \Cref{app:isospin}.

In order to avoid mixing with the three-pion channel, we must restrict the states to have $G=+$. 
This is automatic for the $\pi\pi\eta$ channel,
but not for $K\overline K \pi$, which can have either $G$ parity.
This restriction can be implemented by applying relations between the four $K\overline K \pi$
flavor channels in \Cref{eq:flavorchannels0},
or equivalently upon the corresponding four channels in \Cref{eq:isospindecomp}.
It is simpler to state the restrictions for the latter:\footnote{%
For completeness we note that the restrictions on the channels in \Cref{eq:flavorchannels0}
are as follows: the first and fourth must be 
symmetric under $\bm k_1 \leftrightarrow \bm k_2$,
as must the difference between the second and third channels,
while the sum of the second and third channels must be antisymmetric under
$\bm k_1 \leftrightarrow \bm k_2$.}
To have overall $G=+$, the $K\overline K$ pairs must have $G=-$.
This implies that $[K\bar K]_1$ pairs must be symmetric under $\bm k_1 \leftrightarrow \bm k_2$
(and thus have only even relative angular momentum),
while $[K \bar K]_0$ pairs must be antisymmetric (implying odd relative angular momentum).
This follows because the action on kaons of $G=C e^{-i\pi I_y}$ 
(where $C$ is the charge conjugation operator and $I_y$ the second component of isospin)
is $K^+\to \Kbar$, $K^0\to - K^-$, $\Kbar \to - K^+$, and $K^-\to K^0$.
Here we have used the result that the kaon isodoublets are $(K^+,K^0)$ and $(-\Kbar, K^-)$.
Thus, for example, $[K \bar K]_0\propto K^+ K^- + K^0 \Kbar$,
which under $G$ transforms to $\Kbar K^0 + K^- K^+$. Since the $K$ and $\overline K$
are interchanged, to obtain a negative $G$ parity the relative spatial wavefunction must
be antisymmetric.
We stress that $G$ parity is an exact symmetry of {\em lattice} QCD if $m_u=m_d$,
so that these considerations apply without approximation to the results of simulations.

In the derivation that follows, we will implicitly assume that these restrictions have been applied, so that
only $G=+$ states are present. We will not explicitly apply these restrictions until after the final
form has been obtained. Having the restrictions (implicitly) in place allows us to neglect mixing
with the $3\pi$ channel. We stress that it would be straightforward, though tedious,
to include this channel as a third three-particle state. 
This would result in a separate formalism describing the $3\pi$ and odd-$G$-parity $K\overline K \pi$
system, since $G$ parity is an exact symmetry of isosymmetric QCD.
We choose not to do so in order to keep the focus
on the essential new features that arise when there are additional three-particle channels, and to avoid
cumbersome expressions.

As noted in the introduction, the $b_1(1235)$ and $\eta(1295)$ do not couple to $\pi \pi$ states,
because these resonances have unnatural $J^P$.
However, coupling between $\pi\pi\eta$ or $K\overline K\pi$ and $\pi\pi$ states is possible
for natural $J^P$ other than $0^+$. 
For example, two pions in a $p$-wave combined with an $\eta$ in a relative $p$-wave leads to
an overall $J^P=1^-$, which can mix with two pions.
In finite volume, the rotation group is broken to a finite subgroup,
with parity not being a good symmetry in moving frames.
Thus mixing with two-pion states is allowed in any finite-volume irreducible representations (irreps)
into which natural $J^P$ values are subduced.
Therefore, in the following, we implicitly assume that we are using irreps into which only unnatural
$J^P$ values are subduced, thus avoiding the mixing with two-pion states.
It is straightforward to determine which irreps these are, using, for example, the
group-theoretical results presented in ref.~\cite{Dudek:2012gj}.
For the rest frame, where parity is a good symmetry, we can use the $A_{1u}$, $T_{1g}$,
$A_{2g}$, $E_u$, and $T_{2u}$ irreps, with the first two, respectively, picking out the
$\eta(1295)$ and $b_1(1235)$ quantum numbers.
In other frames, we can only use the $A_2$ irrep, which couples to both
the $\eta(1295)$ and $b_1(1235)$, except for the following caveat.
For frames with momenta of the form $\{0,0,n\}$, $\{0,n,n\}$, $\{n,n,n\}$, $\{n,n,m\}$, and $\{n,m,0\}$,
natural $J^P$ values do appear in the $A_2$ irrep, 
but only starting at $4^+$, $2^+$, $3^-$, $1^-$ and $1^-$, respectively.
Thus we can only use these frames if we assume that the mixing is small in the corresponding
values of $J^P$. This appears most reasonable for the $\{0,0,n\}$ and $\{n,n,n\}$ frames.

\section{Derivation of quantization condition}
\label{sec:topt_derivation}

We follow the TOPT approach introduced in ref.~\cite{\BSQC}, a work hereafter referred to as BS1.
In particular, we make extensive use of the results for three distinguishable particles
(derived in ref.~\cite{\BSnondegen}, referred to as BS3 henceforth)
and for ``2+1'' systems (ref.~\cite{\BStwoplusone}, referred to as BS2).
We use Minkowski time and confine space within a cubic box of side-length $L$ with periodic boundary conditions. 
The starting point is the finite-spatial-volume Minkowski-space correlation matrix,
\begin{equation}
\widehat{C}_L(P)_{jk} \equiv \int d x^0 \int_{L^3} d^3 {\boldsymbol x} \, e^{-i \boldsymbol{P} \cdot {\boldsymbol x}+i E t} \langle 0 \vert \mathrm{T}\mathcal{O}_j(x) \mathcal{O}_k^{\dagger}(0) \vert 0  \rangle_L \,,
\label{eq:CLdef}
\end{equation}
where the indices $i$ and $j$ run over the six flavors given in \Cref{eq:flavorchannels0}, 
with $\mathcal O_j(x)$ being any quasilocal operator possessing the appropriate quantum numbers to annihilate states with flavor $j$, coupling with all allowed irreducible representations of the cubic group. 
The overall momentum vector $\bm P$ is constrained by the boundary conditions to lie in the finite volume set 
$2\pi \bm n/L$, where $\bm n \in \mathbb Z^3$.

For a specified $\bm P$, the energies of finite-volume states are determined by the poles of $C_L$ as a function of $E$. Our derivation will hold in a kinematic range where the only relevant on-shell states are 
$\pi\pi\eta$ and $K \overline K \pi$, the thresholds for which are $M_{\pi\pi\eta}\approx 820$ 
and $M_{KK\pi} \approx 1130\;$MeV, respectively, for physical quark masses.
For the heavier-than-physical quark masses often used in present lattice simulations
of multiparticle states, the two channels will lie closer together, since the pion mass rises more
rapidly with increasing quark mass than those of the kaon or $\eta$.
Since we are restricting ourselves to positive $G$ parity, and to irreps that do not mix with $\pi\pi$ states,
the other potentially relevant channels are $4\pi$, $6\pi$, $8\pi$,
$4\pi+\eta$, $6\pi+\eta$ and $K\overline K \pi\pi$.
As noted in the introduction, we will neglect the coupling to these channels, 
but keep in mind that, in a practical application of our formalism,
this approximation will become increasingly inaccurate as the energy increases.
We expect the range of approximate applicability to be
\begin{equation}
M_{\pi\pi\eta} - 2M_\pi=M_\eta  < E^*=\sqrt{E^2-\bm P^2} < M_{KK\pi} + M_\pi\,.
\end{equation}
Here the lower limit is set either by the presence of the single $\eta$ intermediate state (for the $I=0$ channel),
or, more generally, by the closing of the $\pi\pi$ subchannel at the left-hand cut due to two-pion exchange.

\subsection{All orders expression obtained using TOPT}
\label{eq:TOPT1}

Through a straightforward extension of the work in BS2 and BS3, 
we can express the finite-volume-dependent part of $C_L$ in the relevant kinematic range as a geometric series. 
This series involves TOPT Bethe-Salpeter kernels $B_{2,L}$ and $B_3$ situated between cut factors, $D$, 
that carry the singularities associated with three-particle states:
\begin{equation}
\Delta C_L \equiv C_L - C_\infty^{(0)} = A' i D \frac1{1 - i (B_{2,L}+B_3) i D} A + \mathcal O(e^{-M_\pi L}) ,.
\label{eq:CLresult}
\end{equation}
Here, $A'$ and $A$ represent endcaps, encompassing all diagrams connecting the operators 
$\cO_j$ and $\cO_k^\dagger$ to a three-particle intermediate state, 
while $C_\infty^{(0)}$ is the contribution devoid of three-particle intermediate states. 
All quantities in \Cref{eq:CLresult} are matrices, spanning both flavor space and momentum space. 
Momentum indices are denoted as $\{\bm k\}=\{ \bm k_1, \bm k_2, \bm k_3 \}$, 
with the triplets of finite-volume momenta constrained by $\bm k_1+\bm k_2 + \bm k_3=\bm P$. 
The derivation of \Cref{eq:CLresult} relies on the fact that finite-volume momentum sums of nonsingular 
summands can be replaced by the corresponding infinite-volume integrals, 
with corrections exponentially suppressed in $M_\pi L$~\cite{\KSS}. 
Throughout this derivation, such corrections are assumed negligible, and
will be omitted from subsequent expressions. 
A key consequence of this assumption is that the Bethe-Salpeter kernels, the endcaps,
and $C_\infty^{(0)}$ are infinite-volume quantities, 
aside from a trivial $L$ dependence in $B_{2,L}$ to be described below.

We now describe the remaining quantities present in \Cref{eq:CLresult}, starting with the cut factors. 
These are diagonal in flavor and take the form
\begin{align}
D &= {\rm diag}(D_0, D_0, D_0, D_0, \overline{D}_0, \tfrac12\overline{D}_0)\,,
\label{eq:Ddef}
\end{align}
where $D_0$ and $\overline{D}_0$ are standard TOPT energy denominators,
\begin{align}
D_0 &= \delta_{\{\bm p\} \{\bm k\}} \frac1{L^6} \frac1{8\omega_1 \omega_2 \omega_3}
\frac1{E - \omega_1 - \omega_2 -\omega_3}\,,
\\
\overline{D}_0 &= \delta_{\{\bm p\} \{\bm k\}} \frac1{L^6} \frac1{8\omega'_1 \omega'_2 \omega'_3}
\frac1{E - \omega'_1 - \omega'_2 -\omega'_3}\,.
\end{align}
Here $\omega_i \equiv \omega_{p_{i}} = \sqrt{\bm p_i^2 + M_i^2}$ and 
$\omega'_i \equiv \omega_{p_{i}} = \sqrt{\bm p_{i}^{2} + {M'_{i}}^{2}}$ are single-particle energies,
with $M_{i}$ drawn from the ordered set $\{M_{K}, M_{K}, M_{\pi}\}$, 
and $M'_{i}$ drawn from $\{M_{\pi}, M_{\pi}, M_{\eta}\}$.
The momentum-space Kronecker delta $\delta_{\{\bm p\} \{\bm k\}}$ is shorthand for
\begin{equation}
\delta_{\{\bm p\} \{\bm k\}} = \delta_{\bm p_1 \bm k_1} \delta_{\bm p_2 \bm k_2} \delta_{\bm p_3 \bm k_3}
\,.
\end{equation}
Note that in the final entry of $D$, which corresponds to the 2+1 system $\pi^0\pi^0\eta$,
the factor of $\overline{D}_0$ is accompanied by a symmetry factor of $1/2$.

The appearance of two different cut factors, i.e. $D_0$ and $\overline{D}_0$, which have singularities
at different energies, is the first new feature due to the presence of multiple
three-particle channels. This presents no obstacle to deriving the all-orders result \Cref{eq:CLresult}---
one simply has to keep both types of cut factor explicit.

Next we consider the matrix $B_{2,L}$, which contains two-particle Bethe-Salpeter kernels, denoted,
using calligraphic font, as $\cB_2$. These kernels are defined as sums over all TOPT diagrams 
contributing to two-to-two processes that lack two-particle cuts in the $s$ channel. 
In these diagrams, all momentum sums are replaced by integrals, rendering them infinite-volume quantities. 
Extending the analysis from BS2 and BS3, we arrive at the following block form for the flavor structure,
\begin{equation}
B_{2,L} = \begin{pmatrix}
B_{11} & B_{12} \\
B_{21} & B_{22}
\end{pmatrix}
\end{equation}
where
\begin{equation}
{\small
B_{11} =  \begin{pmatrix}
\substack{\cB_{2,L}(\Kplus \Kbar \leftarrow \Kplus \Kbar) \\
+ \cB_{2,L}(\Kplus \piminus \leftarrow \Kplus \piminus) \\
+ \cB_{2,L}(\Kbar \piminus \leftarrow \Kbar \piminus) }
&
\cB_{2,L}(\Kbar \piminus \leftarrow \Kminus \pizero)
&
\cB_{2,L}(\Kplus \piminus \leftarrow \Kzero \pizero)
&
0
\\
\cB_{2,L}(\Kminus \pizero \leftarrow \Kbar \piminus)
&
\substack{\cB_{2,L}(\Kplus \Kminus \leftarrow \Kplus \Kminus) \\
+ \cB_{2,L}(\Kplus \pizero \leftarrow \Kplus \pizero) \\
+ \cB_{2,L}(\Kminus \pizero \leftarrow \Kminus \pizero) }
&
\cB_{2,L}(\Kplus \Kminus \leftarrow \Kzero \Kbar)
&
\cB_{2,L}(\Kplus \pizero \leftarrow \Kzero \piplus)
\\
\cB_{2,L}(\Kzero \pizero \leftarrow \Kplus \piminus)
&
\cB_{2,L}(\Kzero \Kbar \leftarrow \Kplus \Kminus)
&
\substack{\cB_{2,L}(\Kzero \Kbar \leftarrow \Kzero \Kbar) \\
+ \cB_{2,L}(\Kzero \pizero \leftarrow \Kzero \pizero) \\
+ \cB_{2,L}(\Kbar \pizero \leftarrow \Kbar \pizero) }
&
\cB_{2,L}(\Kbar \pizero \leftarrow \Kminus \piplus)
\\
0
&
\cB_{2,L}(\Kzero \piplus \leftarrow \Kplus \pizero)
&
\cB_{2,L}(\Kminus \piplus \leftarrow \Kbar \pizero)
&
\substack{\cB_{2,L}(\Kzero \Kminus \leftarrow \Kzero \Kminus) \\
+ \cB_{2,L}(\Kzero \piplus \leftarrow \Kzero \piplus) \\
+ \cB_{2,L}(\Kminus \piplus \leftarrow \Kminus \piplus) }
\end{pmatrix}\,,
}
\label{eq:B11}
\end{equation}
\begin{equation}
B_{12} = \begin{pmatrix}
\cB_{2,L}(\Kplus \Kbar \leftarrow \piplus \eta)
&
0
\\
0
&
\cB_{2,L}(\Kplus \Kminus \leftarrow \pizero \eta) S^{D}
\\
0
&
\cB_{2,L}(\Kzero \Kbar \leftarrow  \pizero \eta) S^{D}
\\
\cB_{2,L}(\Kzero \Kminus \leftarrow  \piminus \eta)
&
0
\end{pmatrix}\,,
\label{eq:B12}
\end{equation}
\begin{equation}
B_{21} = \begin{pmatrix}
\cB_{2,L}(\piplus \eta \leftarrow \Kplus \Kbar)
&
0
&
0
&
\cB_{2,L}(\piminus \eta \leftarrow \Kzero \Kminus)
\\
0
&
S^{D} \cB_{2,L}(\pizero \eta \leftarrow \Kplus \Kminus)
&
S^{D} \cB_{2,L}(\pizero \eta \leftarrow \Kzero \Kbar)
&
0
\end{pmatrix}\,,
\label{eq:B21}
\end{equation}
and
\begin{equation}
B_{22} = \begin{pmatrix}
\substack{\cB_{2,L}(\piplus \eta \leftarrow \piplus \eta) \\
+ \cB_{2,L}(\piminus \eta \leftarrow \piminus \eta) \\
+ \cB_{2,L}(\piplus \piminus \leftarrow \piplus \piminus) }
&
\cB_{2,L}(\piplus \piminus \leftarrow \pizero \pizero)
\\
\cB_{2,L}(\pizero \pizero \leftarrow \piplus \piminus)
&
\substack{S^{D} \cB_{2,L}(\pizero \eta \leftarrow \pizero \eta) S^{D} \\
+ \cB_{2,L}(\pizero \pizero \leftarrow \pizero \pizero) }
\end{pmatrix}\,.
\label{eq:B22}
\end{equation}
The texture of nonzero entries is dictated by the two-particle interactions capable of 
inducing the necessary changes in flavor composition. For instance, the top-right entry in $B_{12}$
vanishes because there are no two-particle interactions that connect 
$\Kplus \Kbar \piminus$ to $\pizero \pizero \eta$,
since all three particles must change. 

The form of the individual entries in $B_{2,L}$ is exemplified by
\begin{equation}
\cB_{2,L}(\pizero \eta \leftarrow \Kplus \Kminus)_{\{\bm p\}, \{\bm k\}}
 = 2 \omega_{p_2} L^3 \delta_{\bm p_2 \bm k_3} 
 \cB_2[\pizero(\bm p_1) \eta(\bm p_3) \leftarrow \Kplus(\bm k_1) \Kbar(\bm k_2)]\,,
 \end{equation}
which appears in the lower row of $B_{21}$.
The explicit $L$ dependence arises from keeping track of TOPT propagator factors.
The single momentum Kronecker-delta arises because one of the particles spectates
 (here, one of the $\pizero$s).
Note that we have chosen the $\pizero$ particle in the final state of the scattering
to be that with momentum $\bm p_1$.
The other possible choice, $\bm p_2$, is included by the factor of $S^D$, a symmetrization operator defined as
\begin{equation}
S^D = 1 + P^D\,,\qquad
1 = \delta_{\{\bm p\}, \{ \bm k\}}\,,
\qquad
P^D = \delta_{\bm p_1 \bm k_2} \delta_{\bm p_2 \bm k_1} \delta_{\bm p_3 \bm k_3}\,.
\label{eq:SDdef}
\end{equation}
In words, $P^D$ interchanges the first two momenta. 
All other entries in $B_{2,L}$ have a similar form, with the choice of spectator momenta
depending on the flavors involved in the two-particle scattering, and
factors of $S^D$ appearing when a single $\pizero$ is involved in the scattering.

To complete the description, it remains to define $B_3$. The entries in this flavor matrix are the sum over all TOPT diagrams connecting initial and final flavors with no three-particle cuts. Momentum sums are replaced by integrals, rendering all entries infinite-volume quantities. The only properties of $B_3$ that we will need are that it conserves isospin and is symmetric under interchanges of identical particles in either the initial or final states.

\subsection{On-shell projection}
\label{app:symmon}

The next step is to project the kernels and endcaps on either side of each cut factor in \Cref{eq:CLresult}
on shell, using the approach of BS1. 
The new feature due to having two three-particle channels is that the kinematic details of
on-shell projection differ for the two types of cut. This is straightforward to implement.

As explained most extensively in BS2, a convenient first step is to relocate
the symmetrization operators from
the Bethe-Salpeter kernels into the cut factors $D$. 
This can be accomplished by following the same steps as in BS2, 
leveraging the symmetry under exchange of two identical particles. 
We first pull out the factors of $S_D$ in $B_{2,L}$ by writing
\begin{equation}
B_{2,L} = \tilde S_{D} \tilde B_{2,L} \tilde S_{D}\,,
\label{eq:symm2}
\end{equation}
where
\begin{equation}
\tilde S_D \equiv {\rm diag}(1, 1, 1, 1, 1, S^D)\,.
\label{eq:tSDdef}
\end{equation}
 $\tilde B_{2,L}$ contains no factors of $S_D$ and will be given explicitly below.
Next, we introduce a rescaling matrix
\begin{equation}
R= {\rm diag} (1, 1, 1, 1, 1, \tfrac12)\,,
\label{eq:Rdef}
\end{equation}
such that the following identities hold
\begin{equation}
A =\tilde S_{D} R A\,,\ \
A' = A' R \tilde S_{D}\,, \ \
B_{3} = \tilde S_{D} R B_3 R \tilde S_{D}\,.
\end{equation}
These follow from the invariance under the interchange of the two $\pi^0$s,
with the factor of $1/2$ in the last slot of $R$ cancelling the double counting introduced by applying $S_D$.
Then we can rearrange the volume-dependent part of the correlator into the form
\begin{equation}
\Delta C_L =  \tilde A' i D_S \frac1{1 - i (\tilde B_{2,L}+ \tilde B_3) i D_S} \tilde A\,,
\label{eq:CLresulttilde}
\end{equation}
where
\begin{align}
D_{S} &= \tilde S_{D} D \tilde S_{D} 
= {\rm diag} (D_0, D_0, D_0, D_0, \overline{D}_0, \tfrac12 S^{D} \overline{D}_0 S^{D})\,,
\label{eq:symm1}
\end{align}
and
\begin{equation}
\tilde B_{3} = R B_{3} R \,, \quad \tilde A' = A' R \,,\quad \tilde A = R A \,.
\end{equation}

The explicit form of $\tilde B_{2,L}$ is
\begin{equation}
\tilde B_{2,L} = \begin{pmatrix}
B_{11} & \tilde B_{12} \\
\tilde B_{21} & \tilde B_{22}
\end{pmatrix}\,,
\end{equation}
where the top-left block is unchanged from $B_{2,L}$, \Cref{eq:B11},
the off-diagonal blocks $\tilde B_{12}$ and $\tilde B_{21}$ are given by
\Cref{eq:B12} and \Cref{eq:B21}, respectively, except with the factors of $S_D$ removed,
and
\begin{equation}
\tilde B_{22} = \begin{pmatrix}
\substack{\cB_{2,L}(\piplus \eta \leftarrow \piplus \eta) \\
+ \cB_{2,L}(\piminus \eta \leftarrow \piminus \eta) \\
+ \cB_{2,L}(\piplus \piminus \leftarrow \piplus \piminus) }
&
\tfrac12 \cB_{2,L}(\piplus \piminus \leftarrow \pizero \pizero)
\\
\tfrac12 \cB_{2,L}(\pizero \pizero \leftarrow \piplus \piminus)
&
\substack{
\cB_{2,L}(\pizero \eta \leftarrow \pizero \eta)  \\
+ \tfrac14 \cB_{2,L}(\pizero \pizero \leftarrow \pizero \pizero) }
\end{pmatrix}\,.
\label{eq:tB22}
\end{equation}

With the form \Cref{eq:CLresulttilde} in hand, we can now apply the on-shell projection and resummation
methodology of BS1. 
The on-shell projection changes the momentum indices into the standard $\{k\ell m\}$ coordinates of
the three-particle formalism. This involves picking a spectator particle, whose momentum is $\bm k$
(chosen from the finite-volume set), and then projecting the remaining pair onto spherical harmonics
in their rest frame (with components labeled by $\ell m$).
As detailed in BS3, for channels with three distinct particles, 
and in particular the $K \overline K \pi$ channels,
this leads to an enlargement in the flavor space by a factor of three,
corresponding to the three possible choices of spectator.
For the $2+1$ channel $\pi^0\pi^0\eta$, the enlargement factor is two, corresponding to the
two choices of spectator, as explained in BS2.
The only exception to this counting is the $\pi^+\pi^-\eta$ channel,
where it is preferable to enlarge by four rather than three
by decomposing the $\pi^+ \pi^-$ pair into parts with even and odd relative partial waves.
This facilitates the decomposition into isospin channels,
and borrows from the work of ref.~\cite{\tetraquark} on the doubly-charmed tetraquark.

The result of these enlargements is a flavor matrix of dimension $3+3+3+3+4+2=18$.
Each index corresponds to a choice of spectator together with a choice of ``primary'' member of
the pair, i.e. that with respect to which the spherical harmonic decomposition is defined.
The ordering we use is
\begin{align}
\bigg\{ &[\Kplus \Kbar] \piminus,
\ [\Kplus \piminus] \Kbar,
\ [\Kbar \piminus] \Kplus,
\ [\Kplus \Kminus] \pizero,
\ [\Kplus \pizero] \Kminus,
\ [\Kminus \pizero] \Kplus, \nonumber \\
\ &[\Kzero \Kbar] \pizero,
\ [\Kzero \pizero] \Kbar,
\ [\Kbar \pizero] \Kzero,
\ [\Kzero \Kminus] \piplus,
\ [\Kzero \piplus] \Kminus,
\ [\Kminus \piplus] \Kzero, \nonumber \\
\ &[\piplus \piminus]_e \eta,
\ [\piplus \piminus]_o \eta,
\ [\piplus \eta] \piminus,
\ [\piminus \eta] \piplus,
\ [\pizero \pizero] \eta,
\ [\pizero \eta] \pizero
\label{eq:pairspect0}
\bigg\}\,,
\end{align}
where in each case the pair is enclosed in square brackets, 
with the primary member appearing first,
while the third entry corresponds to the spectator.
The subscripts $e$ and $o$ refer to even and odd partial waves, respectively.
Note that we have adopted the convention that the primary member of the pair is defined
with a priority order given by $\{K, \overline K, \pi^\pm, \pi^0, \eta\}$.

On-shell projection converts cut factors into either $F$ or $G$ cuts, depending on whether
the spectator is unchanged or not. 
We adhere to the prescriptions presented in BS2 and BS3, slightly generalized to handle flavor off-diagonal terms following ref.~\cite{\tetraquark}, and obtain
\begin{equation}
\Delta C_L =  \hat A' i \widehat F_G \frac1{1 - i (\widehat{\cK}_{2,L}+ \widehat{\cK}_{\rm df,3}^{(u,u)}) i \widehat F_G}
\widehat A\,.
\label{eq:CLresulthat}
\end{equation}
We now proceed to explain the elements of this result.

The cut-factor matrix $\widehat F_G$ has block-diagonal form,
\begin{equation}
\widehat F_G = {\rm diag}\left( 
F_G^{1+1+1}, F_G^{1+1+1}, F_G^{1+1+1}, F_G^{1+1+1}, F_G^{4d}, F_G^{2+1} \right)\,,
\label{eq:FG0}
\end{equation}
with entries
\begin{align}
F_G^{1+1+1} &=
\begin{pmatrix}
\tilde F^{\pi} & P_\ell \tilde G^{\pi K} P_\ell & \tilde G^{\pi K} P_\ell
\\
P_\ell \tilde G^{K \pi} P_\ell & \tilde F^{K} & \tilde G^{KK}
\\
P_\ell \tilde G^{K \pi}  & \tilde G^{KK} & \tilde F^{K}
\end{pmatrix}\,.
\label{eq:FG1p1p1}
\\
F_G^{4d} &=
\begin{pmatrix}
P_e \tilde F'^{\eta}P_e & 0 & P_e \tilde G'^{\eta \pi} P_\ell  & P_e \tilde G'^{\eta \pi} P_\ell
\\
0 & P_o \tilde F'^{\eta} P_o & -P_o \tilde G'^{\eta \pi} P_\ell  & P_o \tilde G'^{\eta \pi} P_\ell
\\
P_\ell \tilde G'^{\pi \eta} P_e & -P_\ell \tilde G'^{\pi \eta} P_o & \tilde F'^{\pi} & \tilde G'^{\pi \pi}
\\
P_\ell \tilde G'^{\pi \eta} P_e & P_\ell \tilde G'^{\pi \eta} P_o & \tilde G'^{\pi \pi} & \tilde F'^{\pi}
\end{pmatrix}\,,
\label{eq:barFG1p1p1}
\\
F_G^{2+1} &= \begin{pmatrix} P_e \tilde F'^\eta P_e & \sqrt2 P_e \tilde G^{\prime\eta \pi} P_\ell
\\ \sqrt2 P_\ell \tilde G^{\prime \pi \eta}  P_e & \tilde F^{\prime\pi} + \tilde G^{\prime\pi \pi}  \end{pmatrix}\,.
\label{eq:FG2p1}
\end{align}
The standard kinematic functions $\tilde F^i$ and $\tilde G^{ij}$ are defined in \Cref{app:kin}.
Those with a prime have the $\pi\pi\eta$ cut, while those without have the $K\overline K \pi$ cut.
The factors of
\begin{equation}
P^{(\ell)}_{p' \ell' m'; p \ell m}= \delta_{\bm p' \bm p} \delta_{\ell' \ell} \delta_{m' m} (-1)^\ell 
\label{eq:Plm}
\end{equation}
are needed to correct for mismatches in the conventions for primary spectator between $\wt F^i$
and $\wt G^{ij}$.
For example, in the upper-right element of $F_G^{1+1+1}$, the final-state spectator is the pion,
and thus in $\tilde G^{\pi K}$ the initial-state $K \pi$ pair is decomposed in harmonics relative to
the pion direction. This conflicts with our standard convention, requiring
a factor of $P_\ell$ to the right (initial-state) side of $\tilde G^{D\pi}$.
The factors of $P_e=(1+P_\ell)/2$ and $P_o=(1-P_\ell)/2$ project, respectively,
onto even and odd partial waves.
Finally, we stress that neither $\tilde F^i$ nor $\tilde G^{ij}$ contain symmetry factors;
the only symmetry factors here are the occurrences of $\sqrt2$ in $F_G^{2+1}$.

Next we discuss $\widehat{\cK}_{2,L}$, which contains the two-particle $K$ matrices,
and arises from summing products of factors of $\tilde B_{2,L}$ sewn together with principal-value-regulated integrals over phase space. An explicit formula can be given by generalizing
results in BS2 and BS3, but will not be displayed as it is not needed in the following.
The structure of $\widehat{\cK}_{2,L}$ is straightforward:
there are nonzero entries whenever two channels have a shared spectator.
For example, the $\{1,1\}$ entry involves $\Kplus\Kbar \leftarrow \Kplus\Kbar$ scattering
with a $\pi^-$ spectator, and has the explicit form
\begin{equation}
\left[\widehat \cK_{2,L}\right]_{1 k'\ell' m',1 k \ell m} =
\delta_{\bm k' \bm k} 2 \omega_{k_\pi} L^3 \delta_{\ell' \ell} \delta_{m' m}
\cK_2^{\ell}(\Kplus\Kbar \leftarrow \Kplus\Kbar)\,.
\label{eq:K2Lmat11}
\end{equation}
The superscript on $\cK_2^\ell$ indicates the partial wave, and this quantity has
an implicit dependence on the relative momentum in the center-of-momentum frame (CMF) of
the scattering pair.
An example of an offdiagonal element is that in the $\{3,6\}$ slot,
\begin{equation}
\left[\widehat \cK_{2,L}\right]_{3 k'\ell' m',6 k \ell m} =
\delta_{\bm k' \bm k}  2\omega_{k_K} L^3 \delta_{\ell' \ell} \delta_{m' m}
\cK_2^{\ell}(\Kbar\piminus \leftarrow \Kminus\pizero)\,.
\label{eq:K2Lmat36}
\end{equation}
The complete set of nonzero entries of $\widehat{\cK}_{2,L}$ is provided in \Cref{app:K2L}.

The matrix $\widehat{\cK}_{\rm df,3}^{(u,u)}$ is algebraically related to the quantities discussed above,
 including $B_3$ in particular, but again we do not display the result as it is not needed in the following.
All that is essential to know about $\widehat{\cK}_{\rm df,3}^{(u,u)}$ is that it is an infinite-volume quantity, devoid of any singularities associated with three-particle intermediate states, 
and with nonzero entries in all elements. The $(u,u)$ superscript indicates that it is an unsymmetrized quantity,
the precise meaning of which we will explain in \Cref{sec:sym} below.

Finally, the relation between the new endcaps $\widehat A'$ and $\widehat A$ 
(which are, respectively, $6\times 18$ and $18\times 6$ matrices)
to the earlier endcaps is known, but not needed in the following.
Indeed, we will not keep detailed track of changes to the endcaps in the subsequent manipulations.

Below, we will need to convert 18-dimensional matrices in the $\{k\ell m\}$ basis (such
as $\widehat{\cK}_{\rm df,3}^{(u,u)}$) into
6-dimensional matrices in the original index space given by \Cref{eq:flavorchannels0}
and labeled by triplets of momenta $\{\bm p \}$.
Thus we need to recombine the different choices of spectator flavor, and  convert
from the $\{k\ell m\}$ basis to the momentum basis.
A convenient notation for formalizing the basis conversion was introduced
in ref.~\cite{\tetraquark}, and we recall this notation in \Cref{app:kin}.
It involves operators $\bcX_{[kab]}^{\boldsymbol \sigma}$, defined in \Cref{eq:XLdef},
that act from the left on objects with $\{k\ell m\}$ indices, as well as the conjugate
operators that act from the right.
In term of this, the conversion from an 18-d matrix $M_{\rm ch}^{(18)}$ to a 6-d matrix $M_{\rm ch}^{(6)}$,
both in the charge basis, is accomplished by conjugation
\begin{equation}
M_{\rm ch}^{(6)} = \CR \circ M_{\rm ch}^{(18)} \circ \CL\,,
\label{eq:18to6}
\end{equation}
where $\CR$ is a $6\times 18$ matrix of operators with block form
\begin{equation}
\CR = \begin{pmatrix}
\bcX^{(3)} & 0 & 0 & 0 & 0 & 0 
\\
0 & \bcX^{(3)} & 0 & 0 & 0 & 0
\\
0 & 0 & \bcX^{(3)} & 0 & 0 & 0
\\
0 & 0 & 0 & \bcX^{(3)} & 0 & 0
\\
0 & 0 & 0 & 0 & \bcX^{(4)} & 0
\\
0 & 0 & 0 & 0 & 0 & S_{D} \bcX^{(2)} 
\end{pmatrix}\,,
\label{eq:CRdef}
\end{equation}
where
\begin{align}
\begin{split}
\bcX^{(3)}& \equiv \left(\XR312,\ \XR213,\ \XR123\right)\,,
\\
\bcX^{(4)} &\equiv \left( \XR312 P_e,\ \XR312 P_o , \XR213,  \XR123 \right) \,,
\\
\bcX^{(2)}  &\equiv \left(\sqrt{\tfrac12} \XR312, \XR123\right) \,.
\end{split}
\label{eq:bradefs}
\end{align}
The choices of superscripts in these vectors determines which momentum is assigned to each particle. The fixed subscript on these objects, $[kab]$, indicates that the first superscript index corresponds to $k$, the spectator; the second to $a$, the preferred partner in the interacting pair; and the third to $b$, the remaining partner. This ordering follows from the choices in \Cref{eq:flavorchannels0} as well as the conventions for momentum
labels in \Cref{eq:pairspect0}.
The $a$ and $b$ that appear here should not be confused with those appearing in \Cref{eq:isospindecomp}, which label distinct isospin channels.
The symmetry factor of $\sqrt{1/2}$ is explained in BS2, and the
appearance of $S_D$ follows from \Cref{eq:symm2}.

We stress, as explained in BS2 and ref.~\cite{\tetraquark},
that this conversion is only precise in the $L\to\infty$ limit;
this, however, will be sufficient for our purposes in the following.

\subsection{Converting to the total isospin basis}
\label{app:converttoI}

We now convert to the total isospin basis.
We choose the following ordering,
\begin{equation}
\begin{split}
\big\{ &[[K \bar K]_1 \pi]_2, [[K \pi]_{3/2} \bar K]_2, [[\bar K \pi]_{3/2} K]_2, 
[[\pi\pi]_2 \eta]_2, [[\pi \eta]_1 \pi]_2,
\\
&[[K \bar K]_1 \pi]_1,  [[K \bar K]_0 \pi]_1,
[[K \pi]_{3/2} \bar K]_1, [[K \pi]_{1/2} \bar K]_1, 
[[\bar K \pi]_{3/2} K]_1, [[\bar K \pi]_{1/2} K]_1,
\\
&[[\pi\pi]_1 \eta]_1, [[\pi \eta]_1 \pi]_1,
\\
&[[K \bar K]_1 \pi]_0, [[K \pi]_{1/2} \bar K]_0, [[\bar K \pi]_{1/2} K]_0, 
[[\pi\pi]_0 \eta]_0, [[\pi \eta]_1 \pi]_0 \big\}
\label{eq:Ichannels18}
\end{split}
\end{equation}
where, as before, the final entry corresponds to the spectator, and the first entry to
the primary particle in the pair. 
The first five elements have $I=2$, the next eight have $I=1$, and the final five have $I=0$.
Since isospin is an exact symmetry in our setup,
we must find that all matrices block diagonalize according to isospin.

The unitary matrix $C^{(18)}_{{\rm ch}\to {\rm iso}}$ that converts between bases is given in \Cref{app:chtoiso}.
We insert $[C^{(18)}_{{\rm ch}\to {\rm iso}}]^{-1} C^{(18)}_{{\rm ch}\to {\rm iso}}=1$
between all matrices in \Cref{eq:CLresulthat},
such that each of these matrices is converted to the isospin basis by conjugation, as in \Cref{eq:chtoiso},
while the endcaps are rotated.
We simplify notation by using the same symbols for all quantities after conversion
to the isospin basis, so that the result for $\Delta C_L$ maintains exactly the form of \Cref{eq:CLresulthat}.

After conjugation, the matrices $\widehat F_G$, \Cref{eq:FG0}, and $\widehat{\cK}_{2,L}$, given
in \Cref{app:K2L},
are indeed found to be block digaonal in isospin.
The explicit forms are given below in the summary section, \Cref{sec:QC3}.
As for $\widehat{\cK}_{\rm df,3}^{(u,u)}$, all we know is that it must be block diagonal, but
is otherwise of general form, aside from certain symmetry constraints.
We postpone discussion of its form until we reach its symmetrized version below.

At this point we observe that $\Delta C_L$ has a pole, corresponding to a finite-volume energy level,
whenever
\begin{equation}
\det \left[1 + (\widehat \cK_{2,L} + \widehat{\cK}_{\rm df,3}^{(u,u)}) \widehat F_G \right] = 0\,.
\label{eq:QC3asym}
\end{equation}
This is the asymmetric form of the three-particle quantization condition, and has the
standard form first given in BS1.
It has potential utility as a bridge to quantization conditions derived in the 
finite-volume unitarity approach~\cite{\MD}, as outlined in ref.~\cite{\BSequiv}.
We note that the presence of two three-particle channels leads to $\widehat F_G$ having
two types of free-particle singularities, and to the presence of additional channels in the
two- and three-particle K matrices.

\subsection{Expressing results in terms of symmetrized quantities}
\label{sec:sym}

The drawback of the quantization condition \Cref{eq:QC3asym} is
that it contains a three-particle K matrix,
$\widehat{\cK}_{\rm df,3}^{(u,u)}$, 
that is asymmetric.
As discussed in detail in BS1, BS2 and BS3, the elements of this matrix
are defined by summing an infinite series of products of Bethe-Salpeter
kernels, such that, if the external kernel is a $B_2$, then the noninteracting particle is always
the spectator. This means that only a subset of the total number of diagrams are being included.
In this section we use the method developed in BS1, and extended in BS2 and BS3,
to ``symmetrize'' the three-particle K matrix, by which we mean combining contributions
such that all diagrams are included.

The method is most straightforwardly implemented on a different finite-volume quantity than $C_L$,
namely $\cM_{23,L,\rm off}$. This is a finite-volume $3\to3$ correlation function
involving external single-particle legs at fixed momenta (drawn from the finite-volume set).
External TOPT propagators are amputated, and final and initial times are sent to $\pm \infty$, respectively.
The subscript ``$23$” indicates that this quantity contains not only fully-connected contributions,
but also those in which one of the three particles spectates while the others interact.
(A more detailed explanation is given in fig.~2 of BS3.)
The subscript ``off” indicates that this is an off-shell amplitude
because the energy $E$ does not, in general, equal the sum of the external on-shell energies.
Using $\cM_{23,L,\rm off}$ also allows
the determination of the integral equations relating the three-particle K matrix to
the infinite-volume three-particle scattering amplitude, $\cM_3$.

Just as for $C_L$, $\cM_{23,L,\rm off}$ is a $6\times 6$ matrix in flavor space, 
with the channels those of \Cref{eq:flavorchannels0}.
It also has momentum indices $\{ \bm k\}$.
Following the arguments in BS2 and BS3, it has the form
\begin{align}
\cM_{23,L,\rm off} &= (B_{2,L} + B_3)\frac1{1 -i D i(B_{2,L}+B_3)}\,,
\\
&= \tilde S_{D} (\tilde B_{2,L} + \tilde B_3)
\frac1{1 -i D_S i(\tilde B_{2,L} + \tilde B_3)} \tilde S_D\,,
\end{align}
where the quantities are the same as those appearing in $C_L$.
In the second step we have moved symmetrization factors onto the $D$s.
We observe that the correlator can be written in terms of $\cM_{23,L,\rm off}$~\cite{\BSQC}
\begin{equation}
\Delta C_L = A' iD_S A + A' iD_S i\cM_{23,L,\rm off} iD_S A\,.
\label{eq:CLM23Loff}
\end{equation}
This implies that poles in $\cM_{23,L,\rm off}$ can also be used to determine the
finite-volume energy levels.

As in the analysis of $C_L$, we next introduce additional matrix indices as above to convert to 18-d matrix forms, 
and then project on shell (so that the subscript ``off'' is dropped).
There then follows a series of steps that are essentially exact copies of those used in BS2, BS3
and ref.~\cite{\tetraquark}, and which we do not reproduce here.
These involve tedious algebra, plus the use of symmetrization identities.
The end result is that
\begin{equation}
\begin{split}
\cM_{23,L} &= \CR \circ \widehat{\cM}_{23,L}^{(u,u)} \circ \CL\,,
\\
 \widehat{\cM}_{23,L}^{(u,u)}  &=
 \widehat{\cM}_{2,L} + \widehat{\cD}_L^{(u,u)} + \widehat{\cM}_{{\rm df},3,L}^{(u,u)\prime}\,,
\end{split}
\label{eq:M23Ldef}
\end{equation}
where
\begin{align}
\widehat{\cM}_{2,L} &= \widehat{\cK}_{2,L} \frac1{1-i \widehat F i \widehat{\cK}_{2,L}} \,.
\label{eq:M2L}
\\
\widehat{\cD}_L^{(u,u)} &=  - \widehat{\cM}_{2,L}\widehat G\widehat{\cM}_{2,L}
\frac1{1 + \widehat{G} \widehat{\cM}_{2,L}}\,,
\label{eq:DLuu}
\\
\widehat{\cM}_{\df,3,L}^{(u,u)\prime} &= 
\left[ \frac13 - \widehat{\cD}_{23,L}^{(u,u)} \widehat F \right]
\widehat{\cK}_{\df,3} \frac1{1 + \widehat F_3 \widehat{\cK}_{\df,3}}
\left[\frac13 - \widehat F \widehat{\cD}_{23,L}^{(u,u)} \right]\,,
\label{eq:Mhatdf3L}
\\
\widehat{\cD}_{23,L}^{(u,u)} &= \widehat{\cM}_{2,L}+ \widehat{\cD}_{L}^{(u,u)}
\label{eq:D23Luu}
\\
\widehat F_3 &= \frac{\widehat F}3 - \widehat F \frac1{(\widehat{\cK}_{2,L})^{-1} + \widehat F_G} 
\widehat F\,.
\label{eq:F3app}
\end{align}
The quantities in these expressions are the same as earlier, with $\CR$ given in \Cref{eq:CRdef}.
We have also used the decomposition
\begin{equation}
\widehat F_G = \widehat F + \widehat G\,,
\end{equation}
where $\widehat F$ and $\widehat G$ contain, respectively,
only the $\wt F$ and $\wt G$ terms contributing to $\widehat F_G$,
whose form is given in \Cref{eq:FG0,eq:FG1p1p1,eq:barFG1p1p1,eq:FG2p1}.

Various features of \Cref{eq:M23Ldef} deserve discussion.
First, it was noted earlier that $\cM_{23,L}$ includes disconnected contributions.
These lead to the $\widehat \cM_{2,L}$ term, which should be subtracted to obtain
the fully connected $\cM_{3,L}$.
Second, the operators $\CR$ and $\CL$ in \Cref{eq:M23Ldef}
serve to symmetrize $\widehat{\cM}_{23,L}^{(u,u)}$ in the sense discussed earlier in this section,
namely adding together all the terms that contribute to the finite-volume amplitude.
However, as noted when these operators were defined, 
their action leads to terms that can be added only in the infinite-volume limit. 
While sufficient for deriving integral equations for $\cM_3$, as we do below,
this is problematic if one wishes to use the expression \Cref{eq:M23Ldef} in finite volume.
This brings us to the third point,
which is that in BS1, BS2 and BS3, building on the work of refs.~\cite{\HSQCa,\HSQCb},
symmetrization operators were introduced that {\em can} be used in finite volume.
The net effect is that an equation of similar form to \Cref{eq:M23Ldef} can be written,
with $\CR$ and $\CL$ replaced by these new operators, 
and effectively inserted into \Cref{eq:CLM23Loff}.
This implies that this properly symmetrized $\cM_{23,L}$ is part of a finite-volume
correlator, and thus its poles determine the finite-volume energies.
Only the third term in $\cM_{23,L}$, namely $\widehat{\cM}_{{\rm df},3,L}^{(u,u)\prime}$, 
can lead to such poles,
since three-particle energies must depend on the three-particle K matrix.
This allows us to read off a new form of the quantization condition,
as is done in the following section.

The fourth feature of \Cref{eq:M23Ldef}  concerns the nature of $\widehat{\cK}_{\df,3}$.
As indicated by the absence of the $(u,u)$ superscript, this is a symmetrized quantity.
This follows from the derivation given in BS1, BS2 and BS3.
In particular, for a given choice of external particles, all contributions are included in its
definition. Different choices of spectator flavors lead to different decompositions into
$\{k\ell m\}$ indices, but the underlying quantity is the same.
This is not the case for $\widehat{\cK}_{\df,3}^{(u,u)}$.
A particularly explicit discussion of this point is given in Appendix A of BS2.

Finally, we note that we are free to rotate from the charge basis to the isospin basis,
leaving the form of all equations unchanged. 
In the following, we assume that this rotation has been carried out.

\subsection{Symmetric form of three-particle quantization condition}
\label{sec:QC3}

As just discussed, 
the poles in $\widehat{\cM}_{{\rm df},3,L}^{(u,u)\prime}$
correspond to finite-volume energy levels.
This implies a second form for the quantization condition,\footnote{%
As shown in \Cref{app:sym}, one can also derive this form directly from
the asymmetric form of the quantization condition.}
\begin{equation}
\det\left[1 + \widehat F_3 \widehat{\cK}_{\df,3} \right] = 0\,,
\label{eq:QC3sym}
\end{equation}
where $\widehat F_3$ is given in \Cref{eq:F3app}.
This result takes the standard form for all quantization conditions obtained previously in the RFT 
approach~\cite{\HSQCa,\BHSQC,\BHSK,\BSnondegen,\BStwoplusone,\isospin,\threeN,\tetraquark}.
We call this the symmetric form, as it contains a symmetrized three-particle K matrix.
Given that this minimizes the number of independent components of $\Kdf$, it is the
simplest form of the quantization condition to implement in practice.

Since all matrices contained in the quantization condition block diagonalize in isospin,
we can solve the condition separately for each block. This leads to our final form
\begin{equation}
\det\left[1 + \widehat F_3^{[I]} \widehat{\cK}^{[I]}_{\df,3} \right] = 0\,,
\label{eq:QC3I}
\end{equation}
for $I=2,1,0$, with
\begin{equation}
\widehat F_3^{[I]} = \frac{\widehat F^{[I]} }3 - \widehat F^{[I]}
\frac1{(\widehat{\cK}^{[I]} _{2,L})^{-1} + \widehat F_G^{[I]} }\widehat F^{[I]} \,.
\label{eq:F3I}
\end{equation}

We collect here the explicit forms for the isospin blocks, beginning with those for $I=2$,
\begin{align}
\wh F_G^{[I=2]} &= \begin{pmatrix}
F_G^{1+1+1} & 0 \\
0 & F_G^{2+1} 
\end{pmatrix}
\\
\wh \cK_{2,L}^{[I=2]} &= \begin{pmatrix}
\bcK_{2,L}^{K\bar K,I=1} & 0 & 0 & 0 & \bcK_{2,L}^{\pi\eta\leftrightarrow K\bar K,I=1}
\\
0 & \bcK_{2,L}^{K\pi,I=3/2} & 0 & 0 & 0
\\
0 & 0 & \bcK_{2,L}^{K\pi,I=3/2} & 0 & 0
\\
0 & 0 & 0 &\tfrac12\bcK_{2,L}^{\pi\pi,I=2}& 0 
\\
\bcK_{2,L}^{\pi\eta\leftrightarrow K\bar K,I=1} & 0 & 0 & 0 & \bcK_{2,L}^{\pi\eta,I=1}
\end{pmatrix}
\label{eq:K2LI2}
\end{align}
The subblocks in $F_G^{1+1+1}$ and $F_G^{2+1}$ are
given in \Cref{eq:FG1p1p1,eq:FG2p1}, respectively.
The elements of $\wh \cK_{2,L}^{[I=2]}$ contain the underlying two-particle K matrix along with kinematic factors,
and are defined in \Cref{app:kin}.  
The structure of these matrices follows from the ordering of indices given in \Cref{eq:Ichannels18},
with the $I=2$ part given by
\begin{equation}
\left\{ [[K \bar K]_1 \pi]_2, \ [[K \pi]_{3/2} \bar K]_2,\ [[\bar K \pi]_{3/2} K]_2, \
[[\pi\pi]_2 \eta]_2,\ [[\pi \eta]_1 \pi]_2\right\}\,.
\end{equation}
In particular, the diagonal entries of $\wh \cK_{2,L}^{[I=2]}$ correspond to the scattering of the pair,
with a symmetry factor of $1/2$ for the identical particle case,
while the offdiagonal entry is that for which there is a common spectator particle.

The results for the $I=0$ blocks are similar to those for $I=2$, as expected because of the
similarity of the corresponding indices,
\begin{equation}
\left\{ [[K \bar K]_1 \pi]_0, [[K \pi]_{1/2} \bar K]_0, [[\bar K \pi]_{1/2} K]_0, 
[[\pi\pi]_0 \eta]_0, [[\pi \eta]_1 \pi]_0 \right\}\,.
\end{equation}
We find
\begin{align}
\wh F_G^{[I=0]} &= \begin{pmatrix}
\overline{F}_G^{1+1+1} & 0 \\
0 & F_G^{2+1} 
\end{pmatrix}
\\
\wh \cK_{2,L}^{[I=0]} &= \begin{pmatrix}
\bcK_{2,L}^{K\bar K,I=1} & 0 & 0 & 0 & \bcK_{2,L}^{\pi\eta\leftrightarrow K\bar K,I=1}
\\
0 & \bcK_{2,L}^{K\pi,I=1/2} & 0 & 0 & 0
\\
0 & 0 & \bcK_{2,L}^{K\pi,I=1/2} & 0 & 0
\\
0 & 0 & 0 &\tfrac12\bcK_{2,L}^{\pi\pi,I=0}& 0 
\\
\bcK_{2,L}^{\pi\eta\leftrightarrow K\bar K,I=1} & 0 & 0 & 0 & \bcK_{2,L}^{\pi\eta,I=1}
\end{pmatrix}\,,
\label{eq:K2LI0}
\end{align}
where
\begin{equation}
\overline{F}_G^{1+1+1} =
\begin{pmatrix}
\tilde F^{\pi} & -P_\ell \tilde G^{\pi K} P_\ell & -\tilde G^{\pi K} P_\ell
\\
-P_\ell \tilde G^{K \pi} P_\ell & \tilde F^{K} & \tilde G^{KK}
\\
-P_\ell \tilde G^{K \pi}  & \tilde G^{KK} & \tilde F^{K}
\end{pmatrix}
\end{equation}
differs from $F_G^{1+1+1}$, given in \Cref{eq:FG1p1p1}, by signs in some of the offdiagonal terms.

Finally, for $I=1$, for which the indices are
\begin{multline}
\big\{ [[K \bar K]_1 \pi]_1,  [[K \bar K]_0 \pi]_1,
[[K \pi]_{3/2} \bar K]_1, [[K \pi]_{1/2} \bar K]_1, 
[[\bar K \pi]_{3/2} K]_1, [[\bar K \pi]_{1/2} K]_1,\\
[[\pi\pi]_1 \eta]_1, [[\pi \eta]_1 \pi]_1 \big\}\,,
\end{multline}
we obtain
\begin{align}
\wh F_G^{[I=1]} &= \begin{pmatrix}
F_G^{(6)} & 0 \\
0 & \overline F_G^{2+1} 
\end{pmatrix}
\end{align}
where
\begin{equation}
\overline F_G^{2+1} = \begin{pmatrix} 
P_e \tilde F'^\eta P_e & -\sqrt2 P_e \tilde G'^{\eta \pi} P_\ell
\\ 
-\sqrt2 P_\ell \tilde G'^{\pi \eta}  P_e & \tilde F'^{\pi} - \tilde G'^{\pi \pi}  \end{pmatrix}\,,
\label{eq:FG2p1bar}
\end{equation}
which differs from $F_G^{2+1}$, \Cref{eq:FG2p1}, by several signs,
and
\tiny
\begin{equation}
F_G^{(6)} =
\begin{pmatrix}
\wt F^{\pi} & 0 & -\sqrt{\tfrac13} P_\ell \wt G^{\pi K} P_\ell & \sqrt{\tfrac23} P_\ell \wt G^{\pi K} P_\ell &
-\sqrt{\tfrac13} \wt G^{\pi K} P_\ell  & \sqrt{\tfrac23} \wt G^{\pi K} P_\ell
\\[1em]
0 & \wt F^{\pi} & \sqrt{\tfrac23} P_\ell \wt G^{\pi K} P_\ell & \sqrt{\tfrac13} P_\ell \wt G^{\pi K} P_\ell &
-\sqrt{\tfrac23} \wt G^{\pi K} P_\ell  & -\sqrt{\tfrac13} \wt G^{\pi K} P_\ell
\\[1em]
-\sqrt{\frac13} P_\ell \wt G^{K\pi} P_\ell & \sqrt{\frac23} P_\ell \wt G^{K\pi} P_\ell & \wt F^K & 0 &
- \frac13\wt G^{KK} & - \sqrt{\frac89} \wt G^{KK}
\\[1em]
\sqrt{\frac23} P_\ell \wt G^{K\pi} P_\ell & \sqrt{\frac13} P_\ell \wt G^{K\pi} P_\ell & 0 & \wt F^K &
- \sqrt{\frac89} \wt G^{KK} & \frac13 \wt G^{KK}
\\[1em]
-\sqrt{\frac13} P_\ell \wt G^{K\pi} & -\sqrt{\frac23} P_\ell \wt G^{K\pi} & -\frac13 \wt G^{KK} &
-\sqrt{\frac89} \wt G^{KK} & \wt F^K & 0
\\[1em]
\sqrt{\frac23} P_\ell \wt G^{K\pi} & -\sqrt{\frac13} P_\ell \wt G^{K\pi} & -\sqrt{\frac89} \wt G^{KK} &
{\frac13} \wt G^{KK} & 0 & \wt F^K 
\end{pmatrix}\,.
\end{equation}
\normalsize
The two-particle K matrix is
\begin{align}
\wh \cK_{2,L}^{[I=1]} &= \begin{pmatrix}
\bcK_{2,L}^{K\bar K,I=1} & 0 & 0 & 0 & 0 & 0 & 0 &\bcK_{2,L}^{\pi\eta\leftrightarrow K\bar K,I=1}
\\
0 & \bcK_{2,L}^{K\bar K,I=0} & 0 & 0 & 0 & 0 & 0 & 0
\\
0 & 0 & \bcK_{2,L}^{K\pi,I=3/2} & 0 & 0 & 0 & 0 & 0
\\
0 & 0 & 0 & \bcK_{2,L}^{K\pi,I=1/2} & 0 & 0 & 0 & 0
\\
0 & 0 & 0 & 0 & \bcK_{2,L}^{K\pi,I=3/2} & 0 & 0 & 0
\\
0 & 0 & 0 & 0 & 0 & \bcK_{2,L}^{K\pi,I=1/2} & 0 & 0
\\
0 & 0 & 0 & 0 & 0 & 0 &\tfrac12\bcK_{2,L}^{\pi\pi,I=1}& 0 
\\
\bcK_{2,L}^{\pi\eta\leftrightarrow K\bar K,I=1} & 0 & 0 & 0 & 0 & 0 & 0 & \bcK_{2,L}^{\pi\eta,I=1}
\end{pmatrix}\,,
\label{eq:K2LI1}
\end{align}

The form of the isospin blocks of $\widehat{\cK}_{\rm df,3}$ is more complicated to determine,
and we discuss this in the following section.

With the result in hand, we can study the impact of having multiple (here two) three-particle channels.
We saw above that there are two classes of the kinematic functions $\tilde F$ and $\tilde G$, one with
singularities at $\pi\pi\eta$ free energies, the other with singularities at $K\overline K \pi$ free energies.
Above the $K\overline K \pi$ threshold, these free energies can lie close to each other.
For each total isospin, the sub-blocks of $\wh F_G$ containing these two classes are connected
by the two-particle $\pi\eta \leftrightarrow K\overline K$ scattering, and by elements of $\wh{\cK}_{\rm df,3}$.
This connection can therefore lead to finite-volume states that cannot be thought of as close to
either type of free state, but are instead mixed.
The situation is analogous to that for the two-particle quantization condition when there are multiple channels.
From a practical point of view, the impact is simply that the matrices in the quantization condition become larger.

\subsection{Form of \texorpdfstring{$\Kdf$}{the three-body K-matrix}}
\label{sec:Kdfform}

As already noted, the three-particle K matrix block diagonalizes according to total isospin.
In this section we present the general form of these blocks, $\widehat{\cK}_{\df,3}^{[I]}$.
The analysis follows the methodology introduced in appendix A of ref.~\cite{\tetraquark}.

For each choice of $I$, there are some number of independent channels, as shown in 
\Cref{eq:isospindecomp}: two each for $I=2$ and $0$, and three for $I=1$.
These are the distinct asymptotic states that can scatter into one another.
Assuming PT symmetry, this implies that there are three independent $3\to3$ amplitudes
for $I=2$ and $0$ (two diagonal and one offdiagonal)
and six for $I=1$.
Since the corresponding $\widehat{\cK}_{\df,3}^{[I]}$ matrices have $5\times 6/2=15$ independent entries
for $I=0,2$, and $8\times 9/2=36$ for $I=1$,
there must be many relations between entries.
Determining these relations is the task of this section.

We begin by writing down the form in the charged basis based on results for nondegenerate and
$2+1$ systems, and then rotate to the isospin basis. The block structure of the 18-d matrices is
$3+3+3+3+4+2$, where $3$ indicates the nondegenerate blocks, 
$4$ the extended nondegenerate block with even and odd $\pi\pi$ waves separated,
and $2$ the $2+1$ block,
We label these blocks with an index that runs from $1$ to $6$.
Within each block the entries of $\Kdf$ are given by the same underlying function, 
denoted $\cK_{jk}(\{p\},\{k\})$ with $j,k$ being block indices,
expressed in different coordinates, i.e. with different choices of spectator and primary member of the pair.
PT symmetry relates off-diagonal blocks, 
\begin{equation}
\cK_{jk}(\{p_i\},\{k_i\})= \cK_{kj}(\{k_i\},\{p_i\})\,, \qquad (j\ne k)\,,
\end{equation}
so there are 28 independent underlying functions. These must be related in such a way that in the
isospin basis $\Kdf$ is block diagonal. We will not, however, need (or display) these relations.

The structure within each block is an outer product built from the following three vectors,
\begin{align}
\bcY^{(3)\dagger} &=
\begin{pmatrix}
\YL312,\ & \YL213,\ & \YL123
\end{pmatrix}\,,
\\[5pt]
\bcY^{(4)\dagger} &=
\begin{pmatrix}
P_e \YL312,\ & P_o \YL312,\ & \YL213,\ & \YL123
\end{pmatrix}\,,
\\[5pt]
\bcY^{(2)\dagger} &=
\begin{pmatrix}
\sqrt{\tfrac12}\YL312,\ & \YL123
\end{pmatrix}\,,
\end{align}
and their conjugates
\begin{equation}
\bcY^{(3)} = 
\begin{pmatrix}
\YR312 \\[5pt]
\YR213 \\[5pt]
\YR123 
\end{pmatrix}\,,
\quad
\bcY^{(4)} = 
\begin{pmatrix}
\YR312 P_e \\[5pt]
\YR312 P_o \\[5pt]
\YR213 \\[5pt]
\YR123 
\end{pmatrix}, 
\qquad
\bcY^{(2)} = \begin{pmatrix}
\sqrt{\tfrac12} \YR312
\\[5pt]
\YR123
\end{pmatrix}\,.
\end{equation}
Here we are using the operators $\bcY$ and $\bcY^\dagger$, introduced in ref.~\cite{\tetraquark}, 
which are defined in \Cref{app:kin}.
They convert functions of the triplet of on-shell momenta to the $k\ell m$ basis,
and thus are essentially the inverses of the operators $\bcX$ introduced above.
The permutation associated with the $\bcY$ indicates, in order,  the assignment of momenta
to the spectator, the primary member of the pair, and the remaining member of the pair.
The ordering of these permutations within each of the vectors $\bcY$ is determined by
our choice of momentum labels in \Cref{eq:flavorchannels0} and of the ordering within blocks given
in \Cref{eq:pairspect0}.  The factor of $1/\sqrt2$ is explained in BS2.
We note that in $\bcY^{(2)}$ we can freely replace $\YR123$ with $\YR213$ because this
acts on a function that is symmetric under the interchange of the two neutral pions (whose momenta
are labeled $1$ and $2$).

The outer product form depends only on the dimensions of the block. Thus, for example,
the $\{5,3\}$ block, which has dimensions $4\times 3$, contains
$\bcY^{(4)} \cK_{53} \bcY^{(3)\dagger}$
and similarly in other cases. 

After constructing the $18\times 18$ matrix $\widehat{\K}_{\rm df,3}$ in this fashion, 
we rotate to the isospin basis,
and examine the three isospin blocks in turn. 
For the $I=2$ block we find a sum of four outer products
\begin{equation}
  \widehat{\cK}_{\rm df,3}^{[I=2]} = \sum_{x,y \in \{a,b\}}
    \boldsymbol {\mathcal Y}^{[I=2],x} \circ \cK_{\df,3}^{[I=2],xy}(\{\bm p\},\{\bm k\})
     \circ \boldsymbol {\mathcal Y}^{[I=0/2],y\dagger}
  \,,
  \label{eq:Kdf3I2form}
\end{equation}
where 
\begin{align}
\begin{split}
\bcY^{[I=2],a\dagger} &= \left( \YL312,\ \YL213,\ \YL123,\ 0,\ 0 \right)\,,
\\
\bcY^{[I=2],b\dagger} &= \left( 0,\ 0,\ 0,\ \sqrt{\tfrac12} \YL312,\ \YL213 \right)\,,
\end{split}
\label{eq:Y2ab}
\end{align}  
with the conjugate vectors given similarly.
The superscripts $a,b$ here refer to the two independent states that contribute, respectively
$[[K \bar K]_1 \pi]_2$ and $[[\pi\pi]_2 \eta]_2$. 
Thus, for example, $\cK_{\df,3}^{[I=2],ba}$ corresponds to the K matrix for the process
$\pi\pi\eta \leftarrow K \overline K \pi$.
PT invariance implies that $\cK_{\df,3}^{[I=2],ba}(\{\bm p\},\{\bm k\}) = \cK_{\df,3}^{[I=2],ab}(\{\bm k\},\{\bm p\})$,
so that there are only three underlying K matrices, as claimed above.
Finally, we note that, to obtain these results,
we have used the fact that the $I=2$ $\pi\pi\eta$ state is symmetric under the interchange of the two pions.

The choice of normalization of the $\bcY^{[I=2],x}$ is arbitrary, since any changes can be absorbed
by a redefinition of $\cK_{\rm df,3}^{[I=0],xy}$. The normalization that we choose is explained in
the following section, and in particular is such that \Cref{eq:orthog} holds.

The result for the $I=0$ block takes the same form as for $I=2$, 
\begin{equation}
  \widehat{\cK}_{\rm df,3}^{[I=0]} = \sum_{x,y \in \{a,b\}}
    \boldsymbol {\mathcal Y}^{[I=0],x} \circ \cK_{\df,3}^{[I=0],xy} (\{\bm p\},\{\bm k\})
     \circ \boldsymbol {\mathcal Y}^{[I=0],y\dagger}
  \,,
  \label{eq:Kdf3I0form}
\end{equation}
where
\begin{align}
\begin{split}
\bcY^{[I=0],a\dagger} &= \left( \YL312,\ -\YL213,\ -\YL123,\ 0,\ 0 \right)\,,
\\
\bcY^{[I=0],b\dagger} &= \bcY^{[I=2],b\dagger}\,.
\end{split}
\label{eq:Y0ab}
\end{align}  
In this case $a,b$ refer to $[[K \bar K]_1 \pi]_0$ and $[[\pi\pi]_0\eta]_0$, respectively.
Again, using PT symmetry, there are three independent underlying functions.

Finally, for $I=1$, where the block is eight dimensional and there are three underlying states,
we obtain
\begin{equation}
  \widehat{\cK}_{\rm df,3}^{[I=1]} = \sum_{x,y \in \{a,b,c\}}
    \boldsymbol {\mathcal Y}^{[I=1],x} \circ \cK_{\df,3}^{[I=1],xy}  \circ \boldsymbol {\mathcal Y}^{[I=1],y\dagger}
  \,,
  \label{eq:Kdf3I1form}
\end{equation}
where $a,b,c$ correspond to $[[K\bar K]_1 \pi]_1$, $[[K \bar K]_0 \pi]_1$, and $[[\pi\pi]_1 \eta]_1$,
respectively, and the vectors are now
\begin{align}
\begin{split}
\\
\bcY^{[I=1],a\dagger} &= \left(\YL312,\ 0,\ -\sqrt{\tfrac13} \YL213,\ \sqrt{\tfrac23} \YL213,
\ -\sqrt{\tfrac13}\YL123,\ \sqrt{\tfrac23}\YL123,\ 0\,\ 0 \right)\,,
\\
\bcY^{[I=1],b\dagger} &= \left(0,\ \YL312,\ \sqrt{\tfrac23}\YL213,\ \sqrt{\tfrac13}\YL213,
\ -\sqrt{\tfrac23}\YL123,\ - \sqrt{\tfrac13} \YL123,\ 0,\ 0 \right)\,,
\\
\bcY^{[I=1],c\dagger} &= \left( 0,\ 0,\ 0,\ 0,\ 0,\ 0, \ \sqrt{\tfrac12} \YL312,\  \YL213 \right)\,.
\end{split}
\label{eq:Y1abc}
\end{align}
Using PT symmetry, there are here six underlying functions.

The form of the underlying functions in the above expressions is constrained by symmetries.
This will be discussed in \Cref{sec:Kdfparam} below.

\subsection{Integral equations relating $\Kdf$ to $\cM_3$}
\label{sec:inteqs}

Implementation of the three-particle formalism involves two steps: 
first, determine (constraints on) the three-particle
K matrix, $\widehat {\cK}_{\rm df,3}$ using the quantization condition \Cref{eq:QC3sym} 
to fit to finite-volume energies; and, second, determine the three-particle scattering amplitude
$\cM_3$ by solving the integral equations that relate it to $\widehat {\cK}_{\rm df,3}$~\cite{\HSQCb}.
In this section we describe the required integral equations.

These are obtained by taking an appropriate limit of $\cM_{3,L}$, which, from \Cref{eq:M23Ldef}
and subsequent discussion, is given by
\begin{align}
\cM_{3,L} &= \CR \circ  \widehat{\cD}_L^{(u,u)} \circ  \CL\,.
+ \CR \circ \widehat{\cM}_{{\rm df},3,L}^{(u,u)\prime} \circ  \CL\,.
\label{eq:M3Ldef}
\end{align}
The first term describes a ladder of one-particle exchanges, 
as can be seen from its definition in \Cref{eq:DLuu},
 and does not involve $\widehat {\cK}_{\rm df,3}$.
It contains the divergences that are known to be present in three-particle amplitudes.
The second term is the divergence-free part of $\cM_3$, and is denoted $\cM_{\rm df,3}$.
As can be seen from \Cref{eq:Mhatdf3L}, it vanishes unless $\widehat {\cK}_{\rm df,3}$ is nonzero.

The desired integral equations are obtained by reinserting $i\epsilon$ factors in the numerators
of $\wt F$ and $\wt G$, taking the $L\to\infty$ limit, and then sending $\epsilon\to0$~\cite{\HSQCb}
\begin{equation}
\cM_3 = \lim_{\epsilon\to 0} \lim_{L\to\infty} \cM_{3,L}\,.
\label{eq:inteqs}
\end{equation}
We do not write out the detailed form of the integral equations, as these are essentially the same
as those that have appeared in previous RFT works; see, e.g., refs.~\cite{\HSQCa,\threeN} and BS3.
What is new here are the presence of two types of singularity in the one-particle exchange terms,
due to the presence of two distinct intermediate states, namely $K\overline K \pi$ and $\pi\pi\eta$.
This simply leads to additional matrix indices that are built into the formalism.

The additional feature of the present application is the need to decompose into different
total isospin channels. In particular, the ``core'' of the integral equations resulting from \Cref{eq:inteqs}
(i.e. dropping the factors of $\CR$ and $\CL$)
involves 18-d matrices in flavor space, which, as we have seen above, block-diagonalize 
into 5-d, 8-d, and 5-d blocks corresponding to $I=2,1,0$, respectively.
The final integral equations, by contrast, involve 7-d flavor matrices, 
and the isospin blocks are of size 2, 3, and 2, corresponding to the states listed in \Cref{eq:isospindecomp}.
This is because the integral equations are given in the momentum basis, with no choice of
a spectator or pair required.

Putting this together, what we require is the $7\times 18$ matrix of operators
\begin{equation}
\CR_{{\rm iso18}\to {\rm iso7}} = C_{{\rm ch} \to {\rm iso}}\ \CR \ [ C_{{\rm ch} \to {\rm iso}}^{(18)} ]^{-1}\,,
\end{equation}
where the three matrices on the right-hand side are given in \Cref{eq:chtoiso6}, \Cref{eq:CRdef},
and \Cref{app:chtoiso}, respectively.
An important check on the form of these matrices is that $\CR_{{\rm iso18}\to {\rm iso7}}$
should itself be block diagonal, 
with $2\times 5$ blocks for both $I=2$ and $0$, and a $3\times 8$ block for $I=1$.
We indeed find this to be the case.

The explicit form of the $2\times 5$ block for $I=2$ is
\begin{equation}
\CR_{{\rm iso18}\to {\rm iso7}}^{[I=2]} = 
\begin{pmatrix}
\bcX^{[I=2],a} \\ \bcX^{[I=2],b}
\end{pmatrix}\,,
\label{eq:inteqI2}
\end{equation}
where
\begin{equation}
\begin{split}
\bcX^{[I=2],a} &= \left(\XR312,\ \XR213,\ \XR123,\ 0,\ 0\right)\,,
\\
\bcX^{[I=2],b} &= \left(0,\ 0,\ 0,\ \sqrt{2}\XR312,\ \XR123+\XR213\right)\,.
\end{split}
\label{eq:X2}
\end{equation}
For the $I=0$ block we find
\begin{equation}
\CR_{{\rm iso18}\to {\rm iso7}}^{[I=0]} = 
\begin{pmatrix}
\bcX^{[I=0],a} \\ \bcX^{[I=0],b}
\end{pmatrix}\,,
\label{eq:inteqI0}
\end{equation}
where
\begin{equation}
\begin{split}
\bcX^{[I=0],a} &= \left(\XR312,\ -\XR213,\ -\XR123,\ 0,\ 0\right)\,,
\\
\bcX^{[I=0],b} &= \bcX^{[I=2],b}\,.
\end{split}
\label{eq:X0}
\end{equation}
Finally, for $I=1$, the $3\times 8$ block is
\begin{equation}
\CR_{{\rm iso18}\to {\rm iso7}}^{[I=1]} = 
\begin{pmatrix}
\bcX^{[I=1],a} \\ \bcX^{[I=1],b} \\ \bcX^{[I=1],c}
\end{pmatrix}\,,
\label{eq:inteqI1}
\end{equation}
where
\begin{equation}
\begin{split}
\bcX^{[I=1],a} &= \left(\XR312,\ 0,-\sqrt{\tfrac13}\XR213,\sqrt{\tfrac23}\XR213,
-\sqrt{\tfrac13}\XR123, \sqrt{\tfrac23}\XR123,\ 0,\ 0 \right)\,, 
\\
\bcX^{[I=1],b} &= \left(0, \XR312,\sqrt{\tfrac23}\XR213,\sqrt{\tfrac13}\XR213,
-\sqrt{\tfrac23}\XR123, -\sqrt{\tfrac13}\XR123,\ 0,\ 0 \right)\,, 
\\
\bcX^{[I=1],c} &= \left(0,\ 0,\ 0,\ 0,\ 0,\ 0, \sqrt{2}\XR312, -\XR123+\XR213 \right)\,.
\end{split}
\label{eq:X1}
\end{equation}

To be completely explicit, the final result in the isospin basis is
\begin{equation}
\cM^{[I]}_3 = \lim_{\epsilon\to 0} \lim_{L\to\infty} \CR_{{\rm iso18}\to {\rm iso7}}^{[I]} 
\circ \left( \widehat{\cD}_L^{(u,u),[I]} 
+ \widehat{\cM}_{{\rm df},3,L}^{(u,u)\prime, [I]}  \right) \circ 
\CR_{{\rm iso18}\to {\rm iso7}}^{[I]\dagger} \,,
\label{eq:inteqsiso}
\end{equation}
where the quantities inside the parentheses on the right-hand side are given by
\Cref{eq:DLuu} and \Cref{eq:Mhatdf3L}, respectively,
except that all matrices in these relations are restricted to the corresponding isospin block.
As above, if we drop the $\widehat{\cD}_L^{(u,u)}$ term from the right-hand side,
then we obtain the divergence-free amplitude, $\cM^{[I]}_{\rm df,3}$.
We also stress that the operators $\CR$ and $\CL$ convert the core matrices,
which are in the $\{k\ell m\}$ basis, to the momentum basis $\{k\}$, as is appropriate for
a scattering amplitude.

As a final check, we note that in the formal limit in which we treat $\cM_2$ and $\Kdf$ as small,
we find from \Cref{eq:Mhatdf3L} that
\begin{equation}
\widehat{\cM}_{{\rm df},3,L}^{(u,u)\prime, [I]} = \frac19 \widehat{\cK}_{\rm df,3}^{[I]}
\left[1 + \cO(\cM_2,\Kdf) \right] \,.
\end{equation}
Thus, in this limit,
\begin{align}
\left[\cM^{[I]}_{\rm df,3}\right]_{xy} &\equiv \left[\CR_{{\rm iso18}\to {\rm iso7}}^{[I]}  \circ
\widehat{\cM}_{{\rm df},3,L}^{(u,u)\prime, [I]} \circ
\CR_{{\rm iso18}\to {\rm iso7}}^{[I]\dagger}\right]_{xy}
\\
&\to
\left[\CR_{{\rm iso18}\to {\rm iso7}}^{[I]}  \circ
\frac19 \widehat{\cK}_{\rm df,3}^{[I]}
\circ \CR_{{\rm iso18}\to {\rm iso7}}^{[I]\dagger}\right]_{xy}
\\
&=
\sum_{x,x',y',y} \bcX^{[I],x} \circ
\bcY^{[I],x'} \circ \frac19 \cK_{\rm df,3}^{[I],x'y'}
\circ \bcY^{[I],y' \dagger}
\circ \bcX^{[I],y \dagger}
\\
&=
\cK_{\rm df,3}^{[I],xy}
\,,
\label{eq:MdfisKdf}
\end{align}
where in the penultimate step we have used the outer-product form of $\Kdf$ derived in the previous section.
To do so, we have combined the individual isospin results \Cref{eq:Kdf3I2form,eq:Kdf3I0form,eq:Kdf3I1form}
into a generic notation, where $x,x',y',y$ are summed over $\{a,b\}$ for $I=2,0$ and
over $\{a,b,c\}$ for $I=1$.
In the last step we have used,
\begin{equation}
\bcX^{[I],x} \circ \bcY^{[I],x'} = 3 \delta_{xx'}\,,
\label{eq:orthog}
\end{equation}
which can be shown from the results in
\Cref{eq:Y2ab,eq:Y1abc,eq:X2,eq:X0,eq:X1}
given that $\bcX^{\boldsymbol \sigma}_{[kab]}$, defined in \Cref{eq:XRdef},
is the inverse of $\bcY_{\boldsymbol \sigma}^{[kab]}$, defined in \Cref{eq:YRdef}. 
We stress that \Cref{eq:orthog} holds only when acting on components of $\cK_{\rm df,3}^{[I]}$, 
for it relies on the symmetry or antisymmetry under pion exchange in the $\pi\pi\eta$ components.
The normalization of the $\bcY$ was chosen in the previous section so that the right-hand side of
\Cref{eq:orthog} contains the factor of $3$.
This was done so that $\cM_{\rm df,3}$ and $\Kdf$ have the same normalization in the
weak coupling limit, as shown by \Cref{eq:MdfisKdf}.

\subsection{Projection onto positive $G$ parity}
\label{sec:G}

In this section we discuss how the above-described formalism can be projected onto states of
positive $G$ parity. As described in \Cref{sec:overview}, this is necessary to avoid mixing with three-pion channels,
mixing that we have ignored so far in the development of the formalism.

The projection can be applied to the external operators of \Cref{eq:isospindecomp}, in a manner
already described in \Cref{sec:overview}. The action on this 7-d space is by the projection matrix
\begin{equation}
P_+^{(7)} = {\rm diag} \left(\tfrac12 S_{12},\ 1, \tfrac12 S_{12},\ \tfrac12 A_{12},\ 1,\, \tfrac12 S_{12},\ 1\right)\,,
\end{equation}
where $S_{12}=1+P_{12}$ and $A_{12}=1-P_{12}$, with $P_{12}$ the operator that interchanges the
momenta $\bm k_1$ and $\bm k_2$. 
The action of this projector propagates through the formalism, and leads to a reduction in the size of
the matrices entering the quantization condition.
We stress that this projection commutes with total isospin, so we can consider this reduction block by block.

To determine the projectors on the isospin blocks we need the $G$-parity transformation of the
$K \overline K \pi$ operators in \Cref{eq:flavorchannels0},
\begin{equation}
\begin{split}
\Kplus(\bm k_1) \Kbar(\bm k_2) \piminus (\bm k_3) &\to \Kbar(\bm k_1) \Kplus(\bm k_2) \piminus (\bm k_3) 
\,,\\ 
\Kplus(\bm k_1) \Kminus(\bm k_2) \pizero (\bm k_3) &\to - \Kbar(\bm k_1) \Kzero(\bm k_2) \pizero (\bm k_3)
\,,\\ 
\Kzero(\bm k_1) \Kbar(\bm k_2) \pizero (\bm k_3) &\to - \Kminus(\bm k_1) \Kplus(\bm k_2) \pizero (\bm k_3)
\,,\\ 
\Kzero(\bm k_1) \Kminus(\bm k_2) \piplus (\bm k_3) &\to \Kminus(\bm k_1) \Kzero(\bm k_2) \piplus (\bm k_3) 
\,.
\end{split}
\label{eq:Gops}
\end{equation}
Here we have used the action of $G$ described in \Cref{sec:overview}, as well as the negative
$G$-parity of pions.
For the first and fourth operators $G$ parity simply leads to the interchange $\bm k_1 \leftrightarrow \bm k_2$,
while the second and third operators are themselves interchanged, with an overall sign flip,
as well as $\bm k_1 \leftrightarrow \bm k_2$ interchange.
Note that the $\pi\pi\eta$ operators automatically have positive $G$ parity and thus are unchanged.
We now expand out each of the twelve $K \overline K \pi$ operators in the isospin basis \Cref{eq:Ichannels18}
in terms of underlying operators in \Cref{eq:Gops} (with momentum assignments changed, in general),
and apply the $G$-parity transformation.
The results are
\begin{multline}
[[K\bar K]_{1e/o} \pi]_I \to \pm [[K \bar K]_{1e/o} \pi]_I\,,\quad
[[K\bar K]_{0e/o} \pi]_1 \to \mp [[K \bar K]_{0e/o} \pi]_1\,,
\\
[[K \pi ]_{3/2} \bar K]_I \leftrightarrow [[\bar K \pi]_{3/2} K]_I\,,\quad
[[K \pi ]_{1/2} \bar K]_I \leftrightarrow [[\bar K \pi]_{1/2} K]_I\,,
\end{multline}
where $I$ takes all allowed values, and $e$ and $o$ refer, respectively, 
to projections onto even and odd partial waves.

Thus the projectors onto positive $G$ parity are
\begin{align}
P_G^{[I=2]} &= P_G^{[I=0]} =\begin{pmatrix}
P_e & 0 & 0 & 0 & 0 \\
0 & \tfrac12 & \tfrac12 & 0 & 0 \\
0 & \tfrac12 & \tfrac12 & 0 & 0 \\
0 & 0 & 0 & 1 & 0\\
0 & 0 & 0 & 0 & 1
\end{pmatrix}\,,
\\
P_G^{[I=1]} &= \begin{pmatrix}
P_e & 0 & 0 & 0 & 0 & 0 & 0\\
0 & P_o & 0 & 0 & 0 & 0 & 0\\
0 & 0 & \tfrac12 & 0 & \tfrac12 & 0 & 0 & 0\\
0 & 0 & 0 &\tfrac12 & 0 & \tfrac12 & 0 & 0 \\
0 & 0 & \tfrac12 & 0 & \tfrac12 & 0 & 0 & 0\\
0 & 0 & 0 &\tfrac12 & 0 & \tfrac12 & 0 & 0\\
0 & 0 & 0 & 0 & 0 & 0 & 1 & 0\\
0 & 0 & 0 & 0 & 0 & 0 & 0 & 1
\end{pmatrix}\,.
\end{align}
These project the 5-d blocks for $I=2,0$ down to 4-d,
and project the 8-d blocks for $I=1$ down to 6-d.

The result is that the quantization condition for each isospin takes the same form
as above, \Cref{eq:QC3I,eq:F3I},
but with the matrices $\widehat F_G^{[I]}$, $\widehat \cK_{2,L}$, and $\widehat{\cK}_{\rm df,3}$
replaced by their reduced versions.
We first describe these new versions for $\widehat F_G^{[I]}$ and $\widehat \cK_{2,L}$.
For $I=2$, we find
\begin{align}
\wh F_G^{[I=2]} &\to \begin{pmatrix} 
P_e \tilde F^\pi P_e & \sqrt2 P_e \tilde G^{\pi K} P_\ell & 0 & 0
\\
\sqrt2 P_\ell \tilde G^{K\pi} P_e & \tilde F^K + \tilde G^{KK} & 0 & 0
\\
0 & 0 & P_e \tilde F'^\eta P_e & \sqrt2 P_e \tilde G^{\prime\eta \pi} P_\ell
\\ 
0 & 0 & \sqrt2 P_\ell \tilde G^{\prime \pi \eta}  P_e & \tilde F^{\prime\pi} + \tilde G^{\prime\pi \pi} 
\end{pmatrix}\,,
\\
\wh \cK_{2,L}^{[I=2]} &\to \begin{pmatrix}
P_e \bcK_{2,L}^{K\bar K,I=1} P_e & 0 & 0 & P_e \bcK_{2,L}^{\pi\eta\leftrightarrow K\bar K,I=1}
\\
0 & \bcK_{2,L}^{K\pi,I=3/2} & 0 & 0 
\\
0 & 0 &\tfrac12\bcK_{2,L}^{\pi\pi,I=2}& 0 
\\
\bcK_{2,L}^{\pi\eta\leftrightarrow K\bar K,I=1} P_e & 0 & 0 & \bcK_{2,L}^{\pi\eta,I=1}
\end{pmatrix}\,.
\label{eq:K2LI2G}
\end{align}
Several comments are in order.
First, we note that the upper $2\times2$ block in $\widehat F_G^{[I=2]}$ 
now has a similar structure to the lower such block, which is given by 
the 2+1 form $F_G^{2+1}$.
Thus the projection onto $G=+$ in some sense treats the $K$ and $\overline K$ as identical particles.
Second, we could replace $P_e \wt F^\pi P_e$ with $P_e \wt F^\pi$, due to the properties of
$\wt F$, as explained in appendix A of ref.~\cite{\HSQCa}.
Similarly we do not need a $P_e$ on both sides of the top-left entry in $\widehat{\cK}_{2,L}^{[I=2]}$.
Finally, the factors of $P_e$ acting on the offdiagonal $\pi\eta \leftrightarrow K \overline K$ entries in 
$\widehat{\cK}_{2,L}^{[I=2]}$ can be dropped, since the $G=+$ $\pi\eta$ state 
only couples to the $K\overline K$ state with even partial waves.
In all cases, we keep the factors of $P_e$ to illustrate the action of the projectors.

The results for $I=0$ are similar
\begin{align}
\wh F_G^{[I=0]} &\to \begin{pmatrix} 
P_e \tilde F^\pi P_e & - \sqrt2 P_e \tilde G^{\pi K} P_\ell & 0 & 0
\\
- \sqrt2 P_\ell \tilde G^{K\pi} P_e & \tilde F^K + \tilde G^{KK} & 0 & 0
\\
0 & 0 & P_e \tilde F'^\eta P_e & \sqrt2 P_e \tilde G^{\prime\eta \pi} P_\ell
\\ 
0 & 0 & \sqrt2 P_\ell \tilde G^{\prime \pi \eta}  P_e & \tilde F^{\prime\pi} + \tilde G^{\prime\pi \pi} 
\end{pmatrix}\,,
\label{eq:FGI0G}
\\
\wh \cK_{2,L}^{[I=0]} &\to \begin{pmatrix}
P_e \bcK_{2,L}^{K\bar K,I=1} P_e & 0 & 0 & P_e \bcK_{2,L}^{\pi\eta\leftrightarrow K\bar K,I=1}
\\
0 & \bcK_{2,L}^{K\pi,I=1/2} & 0 & 0 
\\
0 & 0 &\tfrac12\bcK_{2,L}^{\pi\pi,I=0}& 0 
\\
\bcK_{2,L}^{\pi\eta\leftrightarrow K\bar K,I=1} P_e & 0 & 0 & \bcK_{2,L}^{\pi\eta,I=1}
\end{pmatrix}\,,
\label{eq:K2LI0G}
\end{align}
and analogous comments apply.

For $I=1$, we find
\begin{align}
\wh F_G^{[I=1]} &\to \begin{pmatrix}
F_G^{(4)} & 0 \\
0 & \overline F_G^{2+1} 
\end{pmatrix}\,,
\end{align}
where $\overline F_G^{2+1}$ is given in \Cref{eq:FG2p1bar},
and
\begin{equation}
F_G^{(4)} =
\begin{pmatrix}
P_e \wt F^{\pi} P_e & 0 & -\sqrt{\tfrac23} P_e \wt G^{\pi K} P_\ell & \sqrt{\tfrac43} P_e \wt G^{\pi K} P_\ell 
\\[1em]
0 & P_o \wt F^{\pi} P_o & -\sqrt{\tfrac43} P_o \wt G^{\pi K} P_\ell &- \sqrt{\tfrac23} P_o \wt G^{\pi K} P_\ell 
\\[1em]
-\sqrt{\frac23} P_\ell \wt G^{K\pi} P_e & -\sqrt{\frac43} P_\ell \wt G^{K\pi} P_o & \wt F^K- \frac13\wt G^{KK} & - \sqrt{\frac89} \wt G^{KK}
\\[1em]
\sqrt{\frac43} P_\ell \wt G^{K\pi} P_e 
& -\sqrt{\frac23} P_\ell \wt G^{K\pi} P_o & -\sqrt{\frac89} \wt G^{KK} & \wt F^K + \tfrac13 \wt G^{KK}
\end{pmatrix}\,,
\end{equation}
while the two-particle K matrix reduces to
\begin{align}
\wh \cK_{2,L}^{[I=1]} &\to \begin{pmatrix}
P_e \bcK_{2,L}^{K\bar K,I=1} P_e & 0 & 0 & 0 & 0 &P_e \bcK_{2,L}^{\pi\eta\leftrightarrow K\bar K,I=1}
\\
0 & P_o \bcK_{2,L}^{K\bar K,I=0} P_o & 0 & 0 & 0 & 0
\\
0 & 0 & \bcK_{2,L}^{K\pi,I=3/2} & 0 & 0 & 0
\\
0 & 0 & 0 & \bcK_{2,L}^{K\pi,I=1/2} & 0 & 0
\\
0 & 0 & 0 & 0 &\tfrac12\bcK_{2,L}^{\pi\pi,I=1}& 0 
\\
\bcK_{2,L}^{\pi\eta\leftrightarrow K\bar K,I=1} P_e & 0 & 0 & 0 & 0 & \bcK_{2,L}^{\pi\eta,I=1}
\end{pmatrix}\,.
\label{eq:K2LI1G}
\end{align}

Next we describe the reduced forms of $\Kdf$. 
The expressions in terms of sums over outer products, \Cref{eq:Kdf3I2form,eq:Kdf3I0form,eq:Kdf3I1form},
remain valid, but the vectors themselves change to
\begin{align}
\begin{split}
\bcY^{[I=2],a\dagger} &\to \left( \tfrac12(\YL312+\YL321),\sqrt{\tfrac12} (\YL213 +\YL123),\ 0,\ 0 \right)\,,
\\
\bcY^{[I=2],b\dagger} &\to \left(0,\ 0,\ \sqrt{\tfrac12} \YL312,\ \YL213 \right)\,,
\end{split}
\label{eq:Y2abG}
\\
\begin{split}
\bcY^{[I=0],a\dagger} &\to \left( \tfrac12(\YL312+\YL321),-\sqrt{\tfrac12} (\YL213 +\YL123),\ 0,\ 0 \right)\,,
\\
\bcY^{[I=0],b\dagger} &\to \left(0,\ 0,\ \sqrt{\tfrac12} \YL312,\ \YL213 \right)\,,
\end{split}
\label{eq:Y0abG}
\\
\begin{split}
\bcY^{[I=1],a\dagger} &\to \left(\tfrac12(\YL312+\YL321),\ 0,-\sqrt{\tfrac16} (\YL213+\YL123),
\sqrt{\tfrac13} (\YL213+\YL123),\ 0\,\ 0 \right)\,,
\\
\bcY^{[I=1],b\dagger} &\to \left(0,\tfrac12(\YL312-\YL321),\sqrt{\tfrac13}(\YL213-\YL123),
\sqrt{\tfrac13}(\YL213-\YL123),\ 0,\ 0 \right)\,,
\\
\bcY^{[I=1],c\dagger} &\to \left(\ 0,\ 0,\ 0,\ 0, \ \sqrt{\tfrac12} \YL312,\  \YL213 \right)\,.
\end{split}
\label{eq:Y1abcG}
\end{align}

The final change is to the integral equations relating $\Kdf$ to $\cM_3$.
Here again the form of the relations, \Cref{eq:inteqsiso}, remains unchanged,
as does the expressions for the conversion matrices, \Cref{eq:inteqI2,eq:inteqI0,eq:inteqI1}.
The vector of operators entering the latter expressions, however, change to
\begin{align}
\begin{split}
\bcX^{[I=2],a} &\to \left(\tfrac12(\XR312+\XR321),\sqrt{\tfrac12}( \XR213+\XR123),\ 0,\ 0\right)\,,
\\
\bcX^{[I=2],b} &\to  \left(0,\ 0,\ \sqrt{2}\XR312,\ \XR123+\XR213\right)\,,
\end{split}
\label{eq:X2G}
\\
\begin{split}
\bcX^{[I=0],a} &\to \left(\tfrac12(\XR312+\XR321),-\sqrt{\tfrac12}(\XR213+\XR123),\ 0,\ 0\right)\,,
\\
\bcX^{[I=0],b} &\to \left(0,\ 0,\ \sqrt{2}\XR312,\ \XR123+\XR213\right)\,,
\end{split}
\label{eq:X0G}
\\
\begin{split}
\bcX^{[I=1],a} &\to \left(\tfrac12(\XR312+\XR321),\ 0,-\sqrt{\tfrac16}(\XR213+\XR123),
\sqrt{\tfrac13}(\XR213+XR123),\ 0,\ 0 \right)\,, 
\\
\bcX^{[I=1],b} &\to \left(0, \tfrac12(\XR312-\XR321),\sqrt{\tfrac13}(\XR213-\XR123),
\sqrt{\tfrac16}(\XR213-XR123),
\ 0,\ 0 \right)\,, 
\\
\bcX^{[I=1],c} &\to \left(0,\ 0,\ 0,\ 0, \sqrt{2}\XR312, -\XR123+\XR213 \right)\,.
\end{split}
\label{eq:X1G}
\end{align}
The orthogonality relation \Cref{eq:orthog} remains true with these changes.

\section{Reduction to single-channel formalism for $E^* < M_{KK\pi}$}
\label{sec:reduction}

Up to this point, we have had in mind working above both $\pi\pi\eta$ and $K\overline K\pi$ thresholds, i.e.
in the regime where both resonances of interest lie.
In this section we consider the range between the thresholds, 
$ M_{\pi\pi\eta} \lesssim E^* \lesssim M_{KK\pi}$.
Here our formalism remains valid,
but we expect that it can be reduced to that for a single three-particle channel by
``integrating out'' the contributions associated with the $K\overline K\pi$ state.
Our aim is to sketch how this reduction occurs.

The first point to observe is that,
due to the presence of our cutoff functions, the $K\overline K\pi$ channel will be automatically
and smoothly turned off as the energy is reduced.
As described in \Cref{app:kin}, this will occur in two stages, first when the $K\overline K$ channel closes,
and second upon the closure of the $K\pi$ channel.
A similar phenomenon occurs in the $DD\pi$ system, which is where the two-stage turn-off was first
noted~\cite{\tetraquark}.
For physical masses, the two closures occur, respectively, at the three-particle energies
\begin{equation}
E_1^* =M_\pi+ 2 \sqrt{M_K^2-M_\pi^2} \approx 1090 {\rm MeV}\ \ {\rm and}\ \
E_2^*= M_K + \sqrt{M_K^2-M_\pi^2} \approx 970 {\rm MeV}\,.
\end{equation}
These lie about $40$ and $160\;$ MeV below the threshold at $M_{KK\pi} \approx 1130\;$MeV,
and both are well above the lower threshold at $M_{\pi\pi\eta}\approx 820\;$MeV.

The regime we are interested in here is thus $E_2^* <  E^* \lesssim M_{KK\pi}$, so that 
two-channel formalism applies but only one of the channels is kinematically open.
In fact, there is a further constraint on the relevance of the considerations of this section,
namely that we cannot integrate out the $K\overline K\pi$ channel if $\Delta=M_{KK\pi}-E^*$ is too small.
This is because there are finite-volume effects arising from the proximity to on-shell $K\overline K\pi$ states
that scale roughly as $\exp(-\Delta L)$.
These are captured by the full, two-channel quantization condition, but not by the reduced version.
Thus we should use $\Delta\gtrsim M_\pi$, since then the errors caused by reduction to a single-channel form
 are comparable to the exponentially-suppressed terms $\sim \exp(-M_\pi L)$ that the formalism does not control.
 This leaves a very small window $E_2^* < E^* \lesssim M_{KK\pi}-M_\pi$ where the considerations of
 the remainder of this section apply.
 Nevertheless, we think the analysis is of interest both from a purely theoretical point of view,
 and because the window of applicability  could be larger in other multichannel systems.
 
 The strategy we follow is to represent the matrices that enter the quantization condition in block form,
 where the blocks correspond to the $\pi\pi\eta$ and $K\overline K\pi$ channels, respectively.
 Thus, for example, in the $I=0$ case after $G$-parity projection, the $4\times4$ flavor matrices
 such as $F_G^{[I=0]}$ in \Cref{eq:FGI0G} have two $\pi\pi\eta$ entries (the last two) and two
 $K\overline K \pi$ entries (the first two).
 In general, these two blocks will have different dimensions.
We write the block form as
\begin{equation}
M = \begin{pmatrix} M_A & M_B \\ M_C & M_D \end{pmatrix}\,,
\end{equation}
and make repeated use of the following standard results (valid if $M^{-1}$ exists)
\begin{align}
\det M &= \det(M_A - M_B M_D^{-1} M_C)\det(M_D)\,,
\label{eq:schura}
\end{align}
\begin{equation}
\begin{split}
[M^{-1}]_A = \left(M_A - M_B M_D^{-1} M_C \right)^{-1}\,, \qquad
[M^{-1}]_D &= \left(M_D - M_C M_A^{-1} M_B \right)^{-1}\,, 
\\
M_A [M^{-1}]_B = - M_B [M^{-1}]_D\,, \qquad
[M^{-1}]_C M_A &= - [M^{-1}]_D M_C \,.
\end{split}
\label{eq:schurb}
\end{equation}
 
 We first consider the case that $\Kdf=0$, so that the quantization condition can be written
 $\det(F_3^{-1})=0$. Here, and in the remainder of this section, we drop isospin superscripts, since the
 considerations are general. For brevity, we also drop carets.
We are thus looking for energies where an eigenvalue of $F_3$ diverges.
Given the form of $F_3$, \Cref{eq:F3I}, and the fact that $F$ only diverges at noninteracting energies,
we can rewrite the quantization condition as~\cite{\dwave}
\begin{equation}
\det(\cK_{2,L}^{-1} + F_G) = 0\,.
\label{eq:QCnoKdf}
\end{equation}
This is also equiavlent to the asymmetric quantization condition, \Cref{eq:QC3asym},
when $\widehat{\cK}_{\rm df,3}^{(u,u)}=0$.
Our aim is to rewrite this as a single-channel three particle quantization condition.
It will turn out to be easier to initially aim for the asymmetric form, which can be written 
\begin{equation}
\det(\overline{\cK}^{-1} + [F_G]_A) = 0\,,\qquad
\overline{\cK} = \overline{\cK}_{2,L} + \overline{\cK}_{\rm df,3}^{(u,u)}\,,
\label{eq:QCasymred}
\end{equation}
where $\overline{\cK}$ is a reduced matrix living only in the $\pi\pi\eta$ block.
Here we allow for the possibility that integrating out the $K\overline K\pi$ channel
leads to the reappearance of $\Kdf$ (here in asymmetric form).

To proceed, we note that $F_G$ is block diagonal with no $B$ or $C$ components, 
so that,  using \Cref{eq:schura}, the quantization condition \Cref{eq:QCnoKdf} becomes
\begin{align}
\det(r_A + f_A - r_B [r_D + f_D]^{-1} r_C) \det(r_D +f_D) = 0\,,
\label{eq:schurd}
\end{align}
where we have introduced the shorthands $r\equiv\cK_{2,L}^{-1}$ and $f\equiv F_G$. 
We now make the key assumption, namely that the second determinant does not vanish.
This is plausible because $f_D\equiv [F_G]_D$ contains only the $K\overline K \pi$ singularity, 
which is finite in our kinematic regime.
With this assumption, the quantization condition becomes
the vanishing of the first determinant in \Cref{eq:schurd}.
This has the desired form of the reduced quantization condition, \Cref{eq:QCasymred}, 
if we take
\begin{equation}
\overline \cK  = \frac1{r_A -  - r_B [r_D + f_D]^{-1} r_C}\,.
\end{equation}
Using \Cref{eq:schurb}, we find,  after some algebraic effort, that
\begin{equation}
\overline \cK =
 [\cK_{2,L}]_A - [\cK_{2,L}]_B [F_G]_D \frac1{1+[\cK_{2,L}]_D [F_G]_D} [\cK_{2,L}]_C\,.
 \label{eq:Kbarres}
 \end{equation}
To achieve our aim, we must decompose this into the form of the second equality in \Cref{eq:QCasymred},
i.e. a $\overline{\cK}_{2,L}$ part that involves a spectator momentum Kronecker-delta plus an overall
factor of $2\omega L^3$, but is otherwise an infinite-volume quantity,
and a $\overline{\cK}_{\rm df,3}^{(u,u)}$ part that is purely an infinite-volume quantity.
The first part is obtained by keeping only the $F$ terms in \Cref{eq:Kbarres},
\begin{equation}
\overline \cK_{2,L} =
 [\cK_{2,L}]_A - [\cK_{2,L}]_B [F]_D \frac1{1+[\cK_{2,L}]_D [F]_D} [\cK_{2,L}]_C\,.
 \label{eq:Kbar2Lres}
 \end{equation}
 This involves the same spectator (a pion) throughout, and the $2\omega L^3$ factors cancel between
 adjacent factors of $F$ and $\cK_{2,L}$, leaving a single overall such factor.
 Concerning $[F]_D$, one might expect the sum-integral difference in $F$ to vanish, 
 up to exponentially-suppressed corrections,
 since the summand/integrand involves $K\overline K\pi$ denominators and is nonsingular  [see \Cref{eq:Ft}].
In fact, because of our definition of the PV-regulated integral~\cite{\HSQCa}, there is an additional
infinite-volume contribution, proportional to the subthreshold part of $\wt \rho$, \Cref{eq:trho}.
This leads to the following interpretation of $\overline \cK_{2,L}$:
the second term adds back the on-shell, but subthreshold, contributions to $\pi\eta$ scattering
that involve $K\overline K$ intermediate states.
By construction, these contributions are not included in $[\cK_{2,L}]_A$, but are needed in the
full single-channel two-particle K matrix when the $K\overline K\pi$ state is integrated out.

The remainder of $\overline \cK$ involves at least one factor of $[G]_D$, which switches the spectator from
a pion to a $K$ or $\overline K$. In fact, it is easy to see that, after expanding the geometric series
in \Cref{eq:Kbarres}, at least two factors of $[G]_D$ are needed to bring the spectator back to being a pion.
Although an explicit all-orders expression can be given, it is not illuminating.
The important properties are that all three particles are involved in the process,
that the various factors of $2\omega L^3$ [including two for each $G$---see \Cref{eq:Gt}]
cancel aside from an inverse for each internal momentum sum,
and that these momentum sums can be converted to integrals up to exponentially suppressed corrections
(since the summands involve $K\overline K\pi$ intermediate states and are thus nonsingular).
The result is a contribution to $\overline{\cK}_{\rm df,3}^{(u,u)}$, an infinite-volume quantity involving
all three particles. It is asymmetric because only the external pions are spectators.
The interpretation of this contribution to the reduced three-particle kernel is simply that it
adds in contributions involving intermediate $K\overline K\pi$ states
that lead to power-law volume effects above the $K\overline K\pi$ threshold,
due to the singularities in $G$, but now lead only to exponentially-suppressed volume dependence.

We now repeat the argument in the general case, but starting with the asymmetric form of the quantization
condition, written as
 \begin{equation}
\det\left([\cK_{2,L} + {\cK}_{\rm df,3}^{(u,u)}]^{-1} + F_G\right) = 0\,.
\label{eq:QCasymb}
\end{equation}
We follow exactly the same steps as above, leading to the reduced quantization condition  \Cref{eq:QCasymred},
in which $\overline{\cK}$ (denoted with a prime to distinguish it from the earlier form) has the form
\begin{equation}
\overline{\cK}' =
 [\cK_{2,L}+{\cK}_{\rm df,3}^{(u,u)}]_A - 
 [\cK_{2,L} + {\cK}_{\rm df,3}^{(u,u)} ]_B [F_G]_D
  \frac1{1+[\cK_{2,L}+{\cK}_{\rm df,3}^{(u,u)}]_D [F_G]_D} 
  [\cK_{2,L} + {\cK}_{\rm df,3}^{(u,u)}]_C\,.
 \label{eq:Kbarresb}
 \end{equation}
 This can be decomposed as
 \begin{equation}
 \overline{\cK}' = \overline \cK_{2,L} + \overline \cK_{\rm df,3}^{(u,u)} + [{\cK}_{\rm df,3}^{(u,u)}]_A 
 + \overline \cK_{\rm df,3}^{\prime(u,u)}\,,
 \label{eq:Kbarprime}
 \end{equation}
 where the first two terms on the right-hand side are those discussed above in the $\Kdf=0$ case,
 the third term is simply the $A$ block of the asymmetric three-particle K matrix,
 and the final term is the new contribution arising from integrating out $K\overline K \pi$ in the presence
 of a non-zero $\Kdf$.
 An example of a contribution to the final term is
 \begin{equation}
  \overline \cK_{\rm df,3}^{\prime(u,u)} \supset
 -[{\cK}_{\rm df,3}^{(u,u)} ]_B [F_G]_D   [\cK_{2,L}]_C
\end{equation}
As above, it is straightforward to see that, for all contributions,
 the $2\omega L^3$ factors combine to convert sums into
integrals up to exponentially-suppressed volume dependence.
The result is that $\overline \cK_{\rm df,3}^{\prime(u,u)}$ is an infinite volume quantity with the correct
properties to be a contribution to the asymmetric $\Kdf$. 
Thus the last three terms in \Cref{eq:Kbarprime} constitute the renormalized reduced three-particle K matrix.

The final step is to symmetrize the quantization condition. This can be done using the symmetrization
identities given in BS2 and BS3, as already noted in \Cref{sec:sym}. 
However, the method used in those works is based on $\cM_{23,L}$, 
rather than the quantization condition. 
Since here we work directly with the latter, we provide the necessary generalization
in \Cref{app:sym}. The conclusion is that we can convert the asymmetric form of the 
reduced quantization condition to the desired symmetric form,
\begin{align}
\det\left[1 + \overline F_3 \overline{\cK}_{\df,3} \right] &= 0\,,
\label{eq:QC3symred}
\\
\overline F_3 &= \frac{F_A}3 - F_A \frac1{({\overline \cK}_{2,L})^{-1} + [F_G]_A} F_A \,.
\label{eq:F3red}
\end{align}
in which $\overline{\cK}_{\df,3}$ is a symmetrized three-particle
K matrix that is algebraically related to $\overline \cK_{\rm df,3}^{\prime(u,u)}$.

An alternative method of determining the reduced, single-channel quantization condition
is to begin with the full symmetrized quantization condition, expressed as
\begin{equation}
\det\left[{\cK}_{\df,3} + F_{3}^{-1} \right] = 0\,,
\label{eq:QC3symAlt}
\end{equation}
and to apply \Cref{eq:schura} with $M = {\cK}_{\df,3} + F_{3}^{-1}$.
Expanding inside the first determinant of \Cref{eq:schura}, one can identify $[F_{3}^{-1}]_{A}$
and $-[F_{3}^{-1}]_{B} ([F_{3}^{-1}]_{A})^{-1} [F_{3}^{-1}]_{C}$ as the only terms
with overlap of $\overline{F}_{3}^{-1}$, the quantity appearing in the new single-channel quantization condition.
As with the method described above, these terms split into two parts:
those that contribute to $\overline{\cK}_{\df,3}$,
and those which can be folded into $\overline{F}_{3}^{-1}$ through the correct definition of ${\overline \cK}_{2,L}$.
Both methods arrive at the same expression for ${\overline \cK}_{2,L}$, given in \Cref{eq:Kbar2Lres},
and we expect $\overline{\cK}_{\df,3}$ can be brought into the same form through symmetrization.

In summary, we have shown explicitly how, if one goes far enough below the $K\overline K\pi$ threshold,
the quantization condition reduces to the expected form for a single channel system, with a renormalized
$\cK_2$ and $\cK_{\rm df,3}^{(u,u)}$.

\section{Parametrizations of $\Kdf$}
\label{sec:Kdfparam}

In a practical application, one must parametrize the three-particle K-matrix.
This should be done respecting the symmetries of the theory,
in particular Lorentz, $P$, and $T$ invariance, and the exchange symmetries of the three-meson states.
Of particular interest are the behavior near threshold and in the vicinity of a three-particle resonance.
We discuss these two regimes in turn, generalizing the methods introduced in refs.~\cite{\dwave,\isospin}. 

Near threshold, $\Kdf$ can be expanded in terms of kinematic invariants that vanish
at threshold. The procedure for doing so was laid out in ref.~\cite{\dwave} in the case of three identical particles,
and has been subsequently generalized to many systems~\cite{\BStwoplusone,\isospin,\threeN,\tetraquark,\kdfnloall}.
The main new feature here is the presence of two thresholds, 
corresponding to the $\pi\pi\eta$ and $K\overline K\pi$ channels. 
 
From \Cref{sec:Kdfform}, we know that the underlying functions that we have to parametrize
are the $\cK_{\rm df,3}^{[I],xy}$ that appear in \Cref{eq:Kdf3I2form,eq:Kdf3I0form,eq:Kdf3I1form},
where $x,y$ run over the different channels.
Each of these K matrices are Lorentz-invariant functions of the
 three incoming on-shell four-momenta, which here we denote $\{k_i\}$,
  and the corresponding outgoing momenta, denoted $\{k'_i\}$.
 For our two channels, 
we use the labels $i=1,2$ for the two particles that are degenerate,
 while $i=3$ denotes the particle with a distinct mass.
 We denote the initial-state masses as $m_i$
 and the final-state masses as $m'_i$.
 The threshold CMF energies for the two channels are then $M=2 m_1 + m_3$ and
 $M'=2 m'_1+m'_3$.
We call the smaller of these $M_{\rm min}$, and use this to set the scale in the
following.

To parameterize the three-particle amplitude $\Kdf$, we use generalized Mandelstam variables,
\begin{equation}
s \equiv (k_1+k_2+k_3)^2, 
\ \
s_{ij} \equiv (k_i+k_j)^2, 
\ \ s_{ij}' \equiv (k_i'+k_j')^2, 
\ \ 
t_{ij} \equiv (k'_i-k_j)^2\,.
\label{eq:Mandelstams3}
\end{equation}
It is convenient to use the following seventeen dimensionless quantities,
\begin{multline}
\Delta \equiv \frac{s - M^2}{M_{\rm min}^2}\,, \ \
\Delta' \equiv \frac{s - M'^2}{M_{\rm min}^2} 
= \Delta + \frac{M'^2 - M^2}{M_{\rm min}^2}\,, \ \
\\
\Delta_i \equiv \frac{s_{jk} - (m_j+m_k)^2}{M_{\rm min}^2} \,, \ \
\quad \Delta_i' \equiv \frac{s_{jk}' - (m'_j+m'_k)^2}{M_{\rm min}^2}\,, \ \
\widetilde{t}_{ij} \equiv \frac{t_{ij} - (m'_i-m_j)^2}{M_{\rm min}^2}\,,
\label{eq:Deltas}
\end{multline}
where, in the definitions of $\Delta_i$ and $\Delta'_i$, $\{i,j,k\}$ form a cylic permutation of $\{1,2,3\}$.
$\Delta$ and $\Delta_i$ vanish at the initial-state threshold, while $\Delta'$ and $\Delta'_i$ vanish at the
final-state threshold. The $\tilde t_{ij}$ only vanish at threshold if the initial and final channels are the same;
in general they vanish if the final ($i$) and initial particles ($j$) are relatively at rest.
We note that these quantities are well-defined below as well as above the corresponding thresholds.

These seventeen quantities in \Cref{eq:Deltas} are constrained by the 
relation between $\Delta$ and $\Delta'$ given in that equation,
as well as the following eight relations, 
seven of which are independent,\footnote{%
In refs.~\cite{\dwave,\BStwoplusone}, the fact that only seven of these relations were independent was missed. 
This had no impact on the threshold expansions developed in these works. The correct counting was also noted in ref.~\cite{\kdfnloall}.}
\begin{multline}
\sum_{i=1}^3\Delta_{i} = \Delta \,, \quad
\sum_{i=1}^3\Delta_{i}' = \Delta' \,, \quad
\sum_{i=1}^3 \widetilde{t}_{ij} = \Delta_{j} - \Delta + \frac{2 m_j (M'-M)}{M_{\rm min}^2}\,, 
\\
\sum_{i=1}^3 \widetilde{t}_{ji} = \Delta'_{j} - \Delta' +  \frac{2 m'_j (M-M')}{M_{\rm min}^2}\,,
\label{eq:constraints}
\end{multline}
where $j=1,2,3$.
This allows us to express $\Kdf$ in terms of the following nine  variables: $\Delta$, $\Delta_1$, $\Delta_2$, $\Delta'_1$, $\Delta'_2$, $\widetilde{t}_{11}$, $\widetilde{t}_{12}$, $\widetilde{t}_{21}$, and $\widetilde{t}_{22}$.
We develop the threshold expansion treating $\Delta$, $\Delta_i$, $\Delta'_i$, and the $\wt t_{ij}$ as small.
Since the thresholds differ, in practice some of these quantities will be larger than others, and an asymmetric
power-counting is appropriate. As the relative sizes depend on the precise kinematics, which varies according
to the values of the underlying quark masses, we simply ignore this point and work to quadratic order 
in all variables, presenting terms of up to linear order here, and collecting the
quadratic terms in \Cref{app:quadratic}. 

As we have seen in~\cref{sec:overview}, the requirement that the $K \overline{K}$ pairs have negative $G$ parity implies $[K \bar{K}]_1$ must be symmetric under $\bm k_1 \leftrightarrow \bm k_2$, while $[K \bar{K}]_0$ must be antisymmetric under this exchange. 
Similarly, the $[\pi\pi]_{0,2}$ states are symmetric, while $[\pi\pi]_1$ is antisymmetric.
Using these symmetry requirements allows us to write down all allowed
 combinations of the nine kinematic variables in our chosen basis. 

To begin, we consider $I=0$ and $I=2$, each of which have two channels [found in~\cref{eq:isospindecomp}], 
and which have the same symmetry properties. 
While $\mathcal{K}_{\rm df,3}^{[I=0,2],ab}$ and $\mathcal{K}_{\rm df,3}^{[I=0,2],ba}$ are 
symmetric under the 
independent exchanges  $k_1 \leftrightarrow k_2$ and $k'_1 \leftrightarrow k'_2$, 
$\mathcal{K}_{\rm df,3}^{[I=0,2],aa}$ and $\mathcal{K}_{\rm df,3}^{[I=0,2],bb}$ 
are also symmetric under PT transformations.
These symmetries lead to the following result for the diagonal terms,
\begin{multline}
M_{\rm min}^2 \mathcal{K}^{[I=0,2],xx}_\text{df,3}(\{k'\},\{k\}) = \cK^{[I]xx}_0 
+ \cK^{[I]xx}_1 \Delta
+ \cK^{[I]xx}_2 (\Delta_{1}\!+\!\Delta_{2}\!+\!\Delta'_{1}\!+\!\Delta'_{2}) 
\\
+ \cK^{[I]xx}_3 (t_{11}\!+\!t_{12}\!+\!t_{21}\!+\!t_{22})
+ \cO(\Delta^2)\,,
\label{eq:KdfthrI02xx}
\end{multline}
where $x=a$ or $b$, and the coefficients $\cK^{[I]xx}_j$ are real and dimensionless.
For the off-diagonal contributions the lack of PT symmetry leads to one additional term at linear order
\begin{multline}
M_{\rm min}^2\mathcal{K}^{[I=0,2],ab}_\text{df,3}(\{k'\},\{k\}) =
M_{\rm min}^2\mathcal{K}^{[I=0,2],ba}_\text{df,3}(\{k\},\{k'\}) 
= \cK^{[I]ab}_0 + \cK^{[I]ab}_1 \Delta
\\
+ \cK^{[I]ab}_2 (\Delta_{1}\!+\!\Delta_{2}\!) 
+ \cK^{[I]ab}_3 (\!\Delta'_{1}\!+\!\Delta'_{2}) 
+ \K^{[I]ab}_4 (t_{11}\!+\!t_{12}\!+\!t_{21}\!+\!t_{22})
+ \cO(\Delta^2)\,,
\label{eq:KdfthrI02ab}
\end{multline}
where again the coefficients are real and dimensionless.
We note that, in both \Cref{eq:KdfthrI02xx,eq:KdfthrI02ab}, the leading coefficients
lead to a contribution that is independent of momenta and thus isotropic. 
Quadratic terms are given in \Cref{eq:KdfthrI02xxquad,eq:KdfthrI02abquad}.

Now we turn our attention to $I=1$, which contains transitions between three independent states 
[found in~\cref{eq:isospindecomp}]. 
While state $a$ is symmetric under $ k_1 \leftrightarrow k_2$, 
states $b$ and $c$ are antisymmetric under this exchange. 
These exchange symmetries, coupled with the PT invariance of  $\mathcal{K}^{[I=1],xx}_\text{df,3}$, 
lead to the following results.
For $\mathcal{K}^{[I=1],aa}_\text{df,3}$ the result has exactly the same form as that for
$I=0,2$, given in \Cref{eq:KdfthrI02xx}.
For the other diagonal terms, we find
\begin{equation}
\mathcal{K}^{[I=1],yy}_\text{df,3}(\{k'\},\{k\}) = \cK_1^{[1]yy} (t_{11}\!-\!t_{12}\!-\!t_{21}\!+\!t_{22})
+ \cO(\Delta^2)\,,
\label{eq:KdfthrI1yy}
\end{equation}
where $y=b$ or $c$.
For the offdiagonal contributions, we obtain
\begin{multline}
\mathcal{K}^{[I=1],ay}_\text{df,3}(\{k'\},\{k\}) = \mathcal{K}^{[I=1],ya}_\text{df,3}(\{k\},\{k'\}) = 
\\
\cK_1^{[1]ab} (\Delta'_{1}\!-\!\Delta'_{2}\!) + 
\cK_2^{[1]ab} (t_{11}\!-\!t_{12}\!+\!t_{21}\!-\!t_{22}) + \cO(\Delta^2)\,.
\label{eq:KdfthrI1ay}
\end{multline}
with $y=b$ or $c$, and
\begin{equation}
\mathcal{K}^{[I=1],bc}_\text{df,3}(\{k'\},\{k\}) = \mathcal{K}^{[I=1],cb}_\text{df,3}(\{k\},\{k'\}) = 
\cK_1^{[1]bc} (t_{11}\!-\!t_{12}\!-\!t_{21}\!+\!t_{22}) + \cO(\Delta^2)\,.
\label{eq:KdfthrI1bc}
\end{equation}
The antisymmetry of the $b,c$ channels leads to the absence of the isotropic contribution
and a smaller number of linear terms.
Quadratic contributions for $I=1$ K matrices are given in
\Cref{eq:KdfthrI1yyquad,eq:KdfthrI1ayquad,eq:KdfthrI1bcquad}.

We next consider how the expressions given above can be augmented in the $I=0$ and $I=1$ channels 
(in which, respectively, the $\eta(1295)$ and the $b_{1}(1235)$ appear) to incorporate an explicit pole 
that satisfies the relevant symmetries and has a factorizable residue. 
It is plausible that such a pole will be needed to describe a resonance in $\cM_3$~\cite{Garofalo:2022pux}.
We are guided here by the work of ref.~\cite{Hansen:2020zhy} addressing the same issue for the $3\pi$ system,
 in particular for the $J^P=0^-$ resonance, $\pi(1300)$, and the $J^P=1^+$ resonance, $a_1(1260)$.
 We stress that the pole terms we give below are to be added to the threshold expansion forms,
 and that we are presenting only the simplest possible expressions in both cases.
By its very nature, a pole term violates the threshold expansion, and thus one does not have any power counting
to restrict the form of the residues. Thus, higher order terms may be needed in practice.
 
The simplest case is the $I=0$ $\eta(1295)$ resonance. This $J^P=0^-$ state couples to both $\pi\pi\eta$ and
$K\overline K\pi$ in $s$-wave states, and thus no momentum dependence is needed in the residue.
Thus the simplest form to add to $\Kdf$ is
\begin{equation}
\mathcal{K}^{[I=0]xy}_\text{df,3} \supset \frac{v_x v_y}{P^2 - E_{0^-}^2}\,,\qquad
(x,y) \in (a,b)\,,
\label{eq:KdfpoleI0}
\end{equation}
where $\bm v = (v_a, v_b)$ is a real vector describing the relative coupling to the two channels,
with $v_a$ and $v_b$ being coefficients to be determined.
We stress that the pole position $E_{0^-}$ will not equal the $\eta(1295)$ mass, since to connect $\Kdf$ to $\cM_3$
requires solving integral equations, and this will, in general, shift the pole position.

Two additional features arise for the $I=1$, $J^P=1^+$ channel needed for the $b_1(1235)$.
The first is due to the resonance being a vector, and thus having a polarization vector. Summing over this
leads to a factor of
\begin{equation}
\sum_\epsilon \epsilon_\mu \epsilon^*_\nu = g_{\mu\nu}^P \equiv g_{\mu\nu} - P_\mu P_\nu/P^2\,.
\end{equation}
The second new feature is the need to contract the open Lorentz indices with four-vectors having the
appropriate symmetries. The simplest form is
\begin{equation}
\mathcal{K}^{[I=1]xy}_\text{df,3} \supset \frac{1}{P^2 - E_{1^+}^2} g_{\mu\nu}^P w'^\mu_x w^\nu_y\,, \qquad
(x,y) \in (a,b,c)\,,
\label{eq:KdfpoleI1}
\end{equation}
where
\begin{equation}
\begin{split}
\bm w^\nu  &= \left(w_a [k_1^\nu+k_2^\nu], w_b[k_1^\nu-k_2^\nu], w_c [k_1^\nu-k_2^\nu] \right)\,,
\\
\bm w'^\mu  &= \left(w_a [k'^\mu_1+k'^\mu_2], w_b[k'^\mu_1-k'^\mu_2], w_c [k'^\mu_1-k'^\mu_2] \right)\,,
\end{split}
\end{equation}
with $w_x$ being real coefficients to be determined.
We note that an additional allowed contribution to the $w_a$ terms proportional to $P^\nu$ can be dropped
since $g_{\mu\nu}^P P^\nu = 0$.

\section{Conclusions}
\label{sec:conc}

In this work, we have generalized the three-particle formalism to accommodate systems involving multiple
three-particle channels. While we have focused on the two-channel $\pi\pi \eta$ and $K K \pi$ system,
the generalization to additional three-particle channels will involve a simple extension of the work presented here.
Indeed, just as in the two-particle case, the generalization to multiple channels is itself not the most challenging
part of the derivation. Instead, the complications mainly arise
from the need to account for multiple subchannels, and to determine isospin and $G$-parity projections.
These aspects are similar to those arising in the formalism developed for the $DD\pi$ system in 
ref.~\cite{\tetraquark}.
The need for $G$-parity projection is a new feature here, and turns out to reduce the dimensionality of the
matrices appearing in the quantization conditions.

The general structure of the quantization conditions and associated
integral equations is the same as that in all previous applications of the RFT approach.
The presence of two channels simply enlarges the space of spectator flavors.
In our final results, given in \Cref{sec:G}, these spaces are of dimension $4$, $6$, and $4$,
respectively, for $I=0$, $1$, and $2$. As for the $D D \pi$ system studied in ref.~\cite{\tetraquark},
the results for $I=0$ and $I=2$ are nearly identical.

The specific channels on which we have focused allow the application to the $b_1(1235)$ and $\eta(1295)$
resonances, as long as one neglects the coupling to channels with four or more particles.
As discussed in \Cref{sec:overview}, for such an application,
one must use finite-volume irreps in which there are no, or minimal, contributions from $\pi\pi$ states.
There are several such irreps, so we do not expect this to be a significant practical limitation.
We hope that, in the next few years, results from lattice QCD will be available that allow one to
study these resonance. This will likely first be with heavier-than-physical quark masses, which will, in fact,
reduce the problem with neglected channels containing four or more particles.

A three-particle formalism has now been developed that encompasses nearly all systems of interest.
The remaining lacuna is exemplified by the Roper resonance, which decays to $N\pi$ and $N\pi\pi$.
This involves multiple channels with both two and three particles, 
as well as nondegenerate particles with a variety of spins.
Given the recent extension of the formalism to three spin-1/2 particles~\cite{\threeN},
and the by now complete understanding of incorporating nondegenerate 
particles~\cite{\BSnondegen,\BStwoplusone,Pang:2020pkl,\tetraquark},
the main challenge  concerns the combination of two- and three-particle channels.
So far, two approaches have been used: one in which the two-particle channel is incorporated 
explicitly~\cite{\BHSQC}, and the other in which it  is introduced by the presence of a bound state in
a two-particle subchannel of the three-particle system~\cite{\largera,Jackura:2020bsk,Dawid:2023jrj,\tetraquark}.
It remains to be seen which approach, if either, is the best choice for generalization to the
Roper system.\footnote{%
The same comment applies to the generalization of the work presented here to irreps in
which the $\pi\pi$ contribution must be included.}

\acknowledgments

We thank Max Hansen and Fernando Romero-L\'opez  for discussions and comments on the manuscript.
This work is supported in part by the U.S. Department of Energy grant No.~DE-SC0011637. 
This work contributes to the goals of the USDOE ExoHad Topical Collaboration, contract DE-SC0023598.

\appendix

\section{Isospin relations}
\label{app:isospin}

Here we give the relations between the isospin-basis states of \Cref{eq:isospindecomp}
and the flavor-basis states of \Cref{eq:flavorchannels0}.
Doing so, it is important to keep in mind that, while the kaon doublet is $(K^+, K^0)$,
the antikaon doublet contains a sign: $(-\Kbar, K^-)$.
For $I=2$, the results are 
\begin{align}
\begin{split}
I=2,a:&\ [[K \Kbar]_1 \pi]_2 \to \sqrt{\tfrac16} \bigg(
-K^+(k_1) \Kbar(k_2) \pi^-(k_3) +\sqrt2 K^+(k_1) K^-(k_2) \pi^0 (k_3) 
\\ &\quad\quad\quad\quad\quad\quad\quad\quad
- \sqrt2 K^0(k_1) \Kbar (k_2) \pi^0(k_3) + K^0(k_1) K^-(k_2) \pi^+(k_3)  \bigg)\,,
\end{split}
\label{eq:I2a}
\\
I=2,b:&\ [[\pi \pi]_2 \eta]_2 \to \sqrt{\tfrac16} \left( \pi^+(k_1) \pi^-(k_2) + 2 \pi^0(k_1) \pi^0(k_2) + \pi^-(k_1)
\pi^+(k_2) \right) \eta(k_3)\,.
\label{eq:I2b}
\end{align}
For $I=1$ we have
\begin{align}
I=1,a:&\ [[K \Kbar]_1 \pi]_1 \to \sqrt{\frac12} \bigg(
-K^+(k_1) \Kbar(k_2) \pi^-(k_3) - K^0(k_1) K^-(k_2) \pi^+(k_3)\bigg)\,,
\label{eq:I1a}
\\
I=1,b:&\ [[K \Kbar]_0 \pi]_1 \to \sqrt{\frac12} \bigg(
K^+(k_1)  K^-(k_2) \pi^0(k_3) + K^0(k_1) \Kbar(k_2) \pi^0(k_3)\bigg)\,,
\label{eq:I1b}
\\
I=1,c:&\ [[\pi \pi]_1 \eta]_1 \to \sqrt{\frac12} \left( \pi^+(k_1) \pi^-(k_2)  - \pi^-(k_1) \pi^+(k_2) \right) \eta(k_3)\,.
\label{eq:I1c}
\end{align}
Finally, for $I=0$ we have
\begin{align}
\begin{split}
I=0,a:&\ [[K \Kbar]_1 \pi]_0 \to \sqrt{\frac16} \bigg(\!\!
-\!\sqrt2 K^+(k_1) \Kbar(k_2) \pi^-(k_3) - K^+(k_1) K^-(k_2) \pi^0 (k_3) 
\\ &\quad\quad\quad\quad\quad\quad\quad\quad
+ K^0(k_1) \Kbar (k_2) \pi^0(k_3) + \sqrt2 K^0(k_1) K^-(k_2) \pi^+(k_3)  \bigg)\,
\end{split}
\label{eqI0a}
\\
I=0,b:&\ [[\pi \pi]_0 \eta]_0 \to \sqrt{\frac13} \left( \pi^+(k_1) \pi^-(k_2) - \pi^0(k_1) \pi^0(k_2) + \pi^-(k_1)
\pi^+(k_2) \right) \eta(k_3)\,,
\label{eq:I0b}
\end{align}

Thus the conversion from the charged to the isospin basis is accomplished by a $7\times 6$ matrix
\begin{align}
\bm v_{\rm iso} &= C_{\rm ch\to\rm iso} \bm v_{\rm ch}\,,
\\
C_{\rm ch\to\rm iso} &= \sqrt{\frac16}\begin{pmatrix}
-1 & \sqrt2 & -\sqrt2 & 1 & 0 & 0
\\
0 & 0 & 0 & 0 & S_{12} & 2
\\
-\sqrt3 & 0 & 0 & -\sqrt3 & 0 & 0
\\
0 & \sqrt3 & \sqrt3 & 0 & 0 & 0
\\
0 & 0 & 0 & 0 & \sqrt3 A_{12} & 0
\\
-\sqrt2 & -1 & 1 & \sqrt2 & 0 & 0
\\
0 & 0 & 0 & 0 & \sqrt2 S_{12} & -\sqrt2
\end{pmatrix}\,,
\label{eq:chtoiso6}
\end{align}
where $S_{12}=1+P_{12}$ is the symmetrization operator acting on the $\bm k_1$ and $\bm k_2$ 
arguments,
with $P_{12}$ permuting these labels,
while $A_{12}=1-P_{12}$ is the corresponding antisymmetrization operator. 

\section{Kinematic functions}
\label{app:kin}

In this appendix we provide definitions of kinematic quantities and auxiliary operators that enter the 
three-particle formalism. These are taken from previous works in the RFT formalism, with the exception
of the discussion of $\widehat \cK_{2,L}$ in the case of channel mixing.

We begin with the $F$ function for nondegenerate particles,
which appears in the main text starting in \Cref{eq:FG1p1p1}.
There are both primed and unprimed versions, corresponding to $\pi\pi\eta$ and $K \overline K \pi$ triplets.
The latter are given, in the notation of ref.~\cite{\BStwoplusone}, by
\begin{multline}
\left[\wt F^{(i)}\right]_{p' \ell' m';p \ell m} =
\delta_{\bm p' \bm p} \frac{H^{(i)}(\bm p)}{2\omega_{p}^{(i)} L^3}
\left[ \frac1{L^3} \sum_{\bm a}^{\rm UV} - \PV \int^{\rm UV} \frac{d^3 a}{(2\pi)^3} \right]
\\
\times \left[
\frac{\cY_{\ell' m'}^*(\bm a^{*(i,j,p)})}{\big(q_{2,p'}^{*(i)}\big)^{\ell'}}
\frac1{4\omega_{a}^{(j)} \omega_{b}^{(k)}
\big(E\!-\!\omega_{p}^{(i)}\!-\!\omega_{a}^{(j)}\!-\!\omega_{b}^{(k)}\big)}
\frac{\cY_{\ell m}(\bm a^{*(i,j,p)})}{\big(q_{2,p}^{*(i)}\big)^{\ell}}
\right]
\,.
\label{eq:Ft}
\end{multline}
Here $\{i,j,k\}$ are a permutation of $\{K, \overline K, \pi\}$. 
The superscript on $\wt F^{(i)}$ indicates the spectator particle,
while the label $j$ refers to the primary particle of the nonspectator pair, 
with $k$ is the flavor of the third particle.
The ordering conventions are given in \Cref{eq:pairspect0}.

Other quantities appearing in \Cref{eq:Ft} are the on-shell energies, exemplified by
\begin{equation}
\omega_p^{(i)} = \sqrt{M_i^2 + \bm p^2}\,;
\end{equation}
the ``third'' momentum $\bm b = \bm P - \bm p - \bm a$;
and the magnitude of the $(jk)$ pair momentum, assuming three on-shell particles,
\begin{align}
\left[q_{2,p}^{*(i)} \right]^2 &= \frac{\lambda (\sigma_p^{(i)},M_j^2, M_k^2)}{4 \sigma_p^{(i)}}\,,
\\
\sigma_p^{(i)} &= (E - \omega_p^{(i)})^2 - (\bm P - \bm p)^2\,,
\\
\lambda(a,b,c) & = a^2 + b^2 + c^2 - 2 a b - 2 a c - 2 b c\,.
\end{align}
The sum over $\bm a$ runs over the finite-volume set, $\bm a = (2\pi/L) \mathbb{Z}^3$,
and the integral over the pole is regulated by the principal value (PV) prescription,
possibly augmented by the $I_{\rm PV}$ function introduced in ref.~\cite{\largera}.
Both sum and integral are regulated in the same manner in the ultraviolet, with the particular choice
only changing $F$ by exponentially-suppressed effects. 
The momentum $\bm a^{*(i,j,p)}$ is the spatial part of the four-momentum resulting 
from boosting $(\omega_a^{(j)}, \bm a)$ to the center of momentum frame (CMF) of the nonspectator pair,
and thus depends on the spectator momentum $\bm p$, the spectator flavor $i$, 
the mass of the primary member of the pair, $M_j$,
as well as (implicitly) the total four-momentum $(E,\bm P)$.
Note that we use ${}^*$ both to indicate a quantity boosted to a pair CMF,
and complex conjugation. Which usage applies should be clear from the context.
The cutoff function $H^{(i)}(\bm p)$ will be defined below.
Finally,
\begin{equation}
\cY_{\ell m}(\bm r) = \sqrt{4\pi} r^\ell Y_{\ell m}(\hat r)
\end{equation}
are harmonic polynomials.

The $G$ function for nondegenerate particles is given by~\cite{\BStwoplusone}
\begin{equation}
\left[\wt G^{(ij)}\right]_{p \ell' m';r \ell m} =
\frac1{2\omega^{(i)}_{p} L^3}
\frac{\cY_{\ell' m'}^*(\bm r^{*(i,j,p)})}{\big(q_{2,p}^{*(i)}\big)^{\ell'}}
\frac{H^{(i)}(\bm p) H^{(j)}(\bm r)}{b_{ij}^2-m_k^2}
\frac{\cY_{\ell m}(\bm p^{*(j,i,r)})}{\big(q_{2,r}^{*(j)}\big)^{\ell}}
\frac1{2\omega^{(j)}_{r} L^3}\,,
\label{eq:Gt}
\end{equation}
where $i$ and $j$ indicate the flavor of the final and initial spectators, respectively.
The only new notation here is that $b_{ij} = (E - \omega_p^{(i)} - \omega_r^{(j)}, \bm P - \bm p -\bm r)$.

We now return to the cutoff function. 
This is defined, following Refs.~\cite{\HSQCa,\implement,\tetraquark}, as
\begin{align}
H^{(i)}(\bm p) &= J(z^{(i)}(\bm p))\,,\quad
z^{(i)}(\bm p) =  (1 + \epsilon_H) 
\frac{\sigma_p^{(i)} - \sigma_{\rm min}^{(i)}}{\sigma_{\rm th}^{(i)} - \sigma_{\rm min}^{(i)}}\,,
\\
J(z) &= \begin{cases}
0 & z\le 0 \\
\exp(-\tfrac1z \exp[-1/(1-z) ] )   & 0 < z < 1 \\
1 & 1 \le z
\end{cases}\,.
\end{align}
Here $\sigma_{\rm th}^{(i)} = (M_j+M_k)^2$ is the value of $\sigma_p^{(i)}$ at the pair's threshold,
while $\epsilon_H$ is a small positive constant introduced to avoid additional power-law finite-volume 
effects~\cite{\implement}.
$J(z)$ is any function that smoothly interpolates between $0$ and $1$, and we have shown one possible
form. 
The quantity $\sigma_{\rm min}^{(i)}$ is the minimum allowed value of $\sigma_p^{(i)}$,
and should be chosen to avoid any singularities in the two-particle K matrix for the pair.
As discussed in detail in ref.~\cite{\tetraquark}, this singularity is typically due to the nearest
left-hand cut associated with exchange of one or more particles.
Here, for diagonal scattering in all the two-particle channels 
($K \overline K$, $K \pi$, and $\overline K \pi$)
the nearest left-hand cut is due to $t$-channel exchange of two pions.
Avoiding this requires that 
\begin{equation}
\sigma_{\rm min}^{(i)} \ge \left( \sqrt{M_j^2-M_\pi^2} + \sqrt{M_k^2 - M_\pi^2} \right)^2\,.
\label{eq:sigmamin}
\end{equation}
We also enforce that $\sigma_{\rm th}^{(i)} > \sigma_{\rm min}^{(i)}$, so that the cutoff function has
nonvanishing subthreshold support.
Numerically, the minimum values for $\sqrt{\sigma_{\rm min}^{(i)}}$ are
$\approx 950$ and $\approx 475\;$MeV, respectively,
for $i=\pi$,  and $i=\overline K/K$.
These should be compared to the values of $\sqrt{\sigma_{\rm th}}$, which are $\approx 990$ and
$630\;$MeV, respectively.
Thus we learn that the cutoff in the $K\overline K$ channel ($i=\pi$) is much closer to threshold
than that in the $\pi K/\pi \overline K$ channels ($i=\overline K/K$). This observation will play a role in the
discussion of \Cref{sec:reduction}.

The primed quantities $\wt F^{\prime (i)}$ and $\wt G^{\prime (ij)}$ are defined in exactly the same manner,
except that $\{i,j,k\}$ is now a permutation of $\{\pi,\pi,\eta\}$.
Strictly speaking, the cutoff functions should be primed, i.e. $H'^{(i)}(\bm p)$, 
to indicate that a new triplet of particles is being used.
The nearest left-hand cut in both $\pi \pi$ and $\pi \eta$ scattering is due to two-pion exchange so that
the result for $\sigma_{\rm min}^{(i)}$, \Cref{eq:sigmamin}, still holds.
Numerically, the minimum values for $\sqrt{\sigma_{\rm min}^{(i)}}$ are 
$0$ and $\approx 535\;$MeV for $i=\eta$ and $\pi$, respectively,
while those for $\sqrt{\sigma_{\rm th}^{(i)}}$ are $\approx 270$ and $685\;$MeV.

We now turn to the $\bcX$ operators, introduced in ref.~\cite{\tetraquark},
and appearing in the main text starting in \Cref{eq:bradefs}.
These convert from the $\{k\ell m\}$ to the momentum basis.
The operator $\bcX_{[kab]}^{\boldsymbol \sigma}$ acts on a vector $f_{k\ell m}$ as
\begin{equation}
\left[ \bcX_{[kab]}^{\boldsymbol \sigma} \circ f\right]( \{\bm p\} )
= \left[\sum_{\ell m} Y^*_{\ell m} (\hat a^*) f_{k\ell m} \right]_{\bm k\to \bm p_{\sigma_1},
\bm a \to \bm p_{\sigma_2}, \bm b \to \bm p_{\sigma_3}}\,,
\label{eq:XRdef}
\end{equation}
where $\boldsymbol \sigma$ is a permutation of $\{1,2,3\}$.
In words, the sum over $\ell m$ yields a function of $\bm k$ and $\hat a^*$.
The former is then equated to $\bm p_{\sigma_1}$ (the spectator momentum),
while the latter,
when boosted to the CMF of the nonspectator pair,
 is set to the direction of $p_{\sigma_2}$. 
 For three on-shell momenta, this completely determines $p_{\sigma_2}$ and also the final momentum $b=P-k-a$,
which is equated with $p_{\sigma_3}$.
The result is a function of the three on-shell momenta $p_1, p_2, p_3$.
The left-acting version $\boldsymbol{\mathcal X}_{[kab]}^{{\boldsymbol \sigma}\dagger}$ is defined analogously,
\begin{align}
\left[f \circ \boldsymbol{\mathcal X}_{[kab]}^{\boldsymbol \sigma \dagger} \right] (\{ k_i\})
= \left[\sum_{\ell m} f_{k\ell m} Y_{\ell m}(\hat a^*)
\right]_{\bm k\to \bm k_{\sigma_1} , \, \bm a\to \bm k_{\sigma_2}, \, \bm b \to \bm k_{\sigma_3}}
\,.
\label{eq:XLdef}
\end{align}

Next we introduce operators $\boldsymbol{\mathcal Y}^{[kab]}_{\boldsymbol \sigma}$
that have the inverse action of the $\boldsymbol{\mathcal X}_{[kab]}^{\boldsymbol \sigma}$,
again borrowing notation from ref.~\cite{\tetraquark}.
These appear first in the main text in \Cref{sec:Kdfform}.
The $\boldsymbol{\mathcal Y}^{[kab]}_{\boldsymbol \sigma}$ act on functions $g(\{p_i\})$ of three on-shell momenta,
yielding objects that have $\{k\ell m\}$ indices:
\begin{align}
\left[{\boldsymbol {\mathcal Y}}_{{{\boldsymbol \sigma}}}^{[kab]} \circ g \right]_{k\ell m}
&=
\frac{1}{4\pi} \int d\Omega_{a^*} Y_{\ell m}(\hat a^*)
g(\{ p_i \})\bigg|_{p_{\sigma(1)}\to k,\ p_{\sigma(2)}\to a, \ p_{\sigma(3)}\to b} \,,
\label{eq:YRdef}
\end{align}
where $\boldsymbol \sigma$ is again a permutation of $\{1,2,3\}$.
In words, we choose $p_{\sigma_1}$ to be the spectator momentum,
leaving $p_{\sigma_2}$ and $p_{\sigma_3}$ to be the remaining pair.
We boost to the CMF of this pair, and decompose into spherical harmonics, defining
$\hat a^*$ as the direction of $\bm p_{\sigma_2}$ in this frame.
An analogous definition holds for the conjugate operator
$\boldsymbol{\mathcal Y}^{[kab]\dagger}_{\boldsymbol \sigma}$,
which acts from the right and includes the complex-conjugated spherical harmonics.

Finally, we give the explicit form for the two-particle K matrices, 
or, more precisely, for the inverse of these matrices,
since it is the latter that enter the quantization condition [see \Cref{eq:QC3I,eq:F3I}].
We focus on channels having definite isospin, rather than using the charge basis,
since the former enter the final form of the quantization condition, both before
(\Cref{sec:QC3}) and after (\Cref{sec:G}) $G$-parity projection.

As can be seen from the form of the isospin blocks $\widehat{\cK}_{2,L}^{[I]}$,
given in \Cref{eq:K2LI2,eq:K2LI0,eq:K2LI1} before $G$-parity projection
and \Cref{eq:K2LI2G,eq:K2LI0G,eq:K2LI1G}  after,
the only mixing occurs between the $[K\bar K]_1$ and $[\pi\eta]_1$ channels.
For the other channels, taking the inverse is straightforward as the matrix is diagonal in all indices,
and we discuss these cases first.
The general form of the inverse is exemplified by
\begin{equation}
\left[\frac1{\bcK_{2,L}^{K\pi,I=3/2}}\right]_{k'\ell' m',k\ell m} = 
\delta_{\bm k' \bm k} \frac1{2 \omega_{k}^{\bar K} L^3} \delta_{\ell' \ell} \delta_{m' m}
\frac1{\cK_2^{K\pi,I=3/2,(\ell)}(q_{2,k}^{*\bar K})}\,,
\label{eq:K2bar}
\end{equation}
where we recall that the superscript $\bar K$ indicates the spectator flavor, and
\begin{equation}
\frac1{\cK_2^{K\pi,I=3/2,(\ell)}(q_{2,k}^{*\bar K})} = \frac{\eta_{\bar K}}{8 \pi \sigma_k^{ \bar K}}
\left\{ q_{2,k}^{* \bar K } \cot \delta_\ell^{K\pi,I=3/2}(q_{2,k}^{*\bar K})
+ \left|q_{2,k}^{*\bar K}\right| [1 - H'^{\bar K}(\bm k) ]
\right\}\,.
\label{eq:K2ellinv}
\end{equation}
$\eta_{\bar K}$ is a symmetry factor that is unity here and in all other channels except for $\pi\pi$,
where it is $1/2$. Note that we are abusing notation slightly as the right-hand side depends on the
spectator momentum $\bm k$ and flavor (here $i=\overline K$), but we do not show this explicitly on the
left-hand side.
The result \Cref{eq:K2ellinv}
can  be generalized to deal with poles in $\cK_2$ by introducing an $I_{\rm PV}$ function,
matching the corresponding addition to $F$ noted earlier.
The details of this function and its role are described in ref.~\cite{\largera}.

It will be helpful for the generalization to the case involving channel mixing to recall the
argument leading to the form of the right hand side in \Cref{eq:K2ellinv}.
To do so we rewrite \Cref{eq:M2L} as
\begin{equation}
[\widehat{\cM}_{2,L}]^{-1} =  [\widehat{\cK}_{2,L}]^{-1}+ \widehat F\,.
\label{eq:M2Linv}
\end{equation}
We first apply this to channels that only have diagonal scattering, where the inverses are trivial.
We take the $L\to \infty$ limit in the fashion described in ref.~\cite{\HSQCb}, i.e. first introducing
the standard $i\epsilon$ into the poles. A detailed description of this limit using the present notation is given in
sec.~VII of ref.~\cite{\BSnondegen}. Removing common factors, one then finds
\begin{align}
[\cM_2^{(i)(\ell)}(q_{2,k}^{*(i)})]^{-1} &= [\cK_2^{(i)(\ell)}(q_{2,k}^{*(i)})]^{-1} + 
H(\bm k) \wt \rho(q_{2,k}^{*(i)}) \,,
\label{eq:M2inv}
\\
\wt \rho(q_{2,k}^{*(i)}) &=
\frac{\eta_i}{8 \pi \sigma_k^{(i)}} \begin{cases}
-i q_{2,k}^{*(i)} & q_{2,k}^{*(i)2} \ge 0
\\
\left|q_{2,k}^{*(i)}\right| & q_{2,k}^{*(i)2} < 0
\end{cases}
\,.
\label{eq:trho}
\end{align}
Here we are denoting the channel generically by the spectator index $i$, and we stress that
$\cM_2^{(i)(\ell)}$ is the physical infinite-volume two-particle scattering amplitude.
The $H \wt \rho$ term in \Cref{eq:M2inv} comes from the $L\to\infty$ limit of $\wt F$, \Cref{eq:Ft},
in which limit the sum-integral difference vanishes aside from the difference between the PV and
$i\epsilon$ pole prescriptions. This leads to the phase-space factor $\wt \rho$, including its analytic
continuation below threshold, as well as the overall factor of $H$ that comes with $\wt F$.
The unitarity constraint on $\cM_2$ can be resolved by writing
\begin{align}
[\cM_2^{(i)(\ell)}(q_{2,k}^{*(i)})]^{-1} &= \frac{\eta_i}{8 \pi \sigma_k^{(i)}}
q_{2,k}^{*(i)} \cot \delta_\ell^{(i)}(q_{2,k}^{*(i)})
+ \wt \rho(q_{2,k}^{*(i)}) \,,
\label{eq:M2invU}
\end{align}
where the first term is the standard continuum definition of $\cK_2^{-1}$.
Equating the results in \Cref{eq:M2inv} and \Cref{eq:M2invU} leads to the form given above
in \Cref{eq:K2ellinv}.
We note that the modified $\cK_2$ that appears in the RFT formalism agrees with the standard one
above threshold (where $H=1$) and smoothly interpolates to $\cM_2$ below threshold, with equality
holding once $H=0$.

We now extend this discussion to the case of two channel mixing. 
We label the two channels $i=1,2$ for $[K \bar K]_1,[\pi \eta]_1$, respectively.
We use the Blatt-Biedenharn (BB) parametrization of the
S matrix~\cite{PhysRev.86.399},
\begin{equation}
S_2 = 
\begin{pmatrix}
c_\epsilon & - s_\epsilon \\
s_\epsilon & c_\epsilon
\end{pmatrix}
\begin{pmatrix}
e^{2i\delta_1} & 0
\\
0 & e^{2i \delta_2}
\end{pmatrix}
\begin{pmatrix}
c_\epsilon & s_\epsilon \\
-s_\epsilon & c_\epsilon
\end{pmatrix}\,,
\label{eq:S2BB}
\end{equation}
where $c_\epsilon \equiv \cos(\epsilon)$, $s_\epsilon \equiv \sin(\epsilon)$, with $\epsilon$ the mixing angle,
and all quantities have an implicit label $(\ell)$.
The relation to the scattering amplitude is~\cite{Hansen:2012tf}
\begin{equation}
S_2 = 1 +2  i P \cM_2 P\,,\qquad 
P = {\rm diag} \left( \sqrt{\frac{\eta_1 q_1^*}{8 \pi E^*}},\ \sqrt{\frac{\eta_2 q_2^*}{8 \pi E^*}} \right)\,,
\label{eq:M2fromS2}
\end{equation}
where we are using $E^*$ in place of $\sqrt{\sigma_k}$ for the total energy in the CMF, which is common
to both channels, and $q_i^*$ are the channel-dependent CMF particle momenta.
We include symmetry factors to keep the discussion general, although for the channels of interest
$\eta_1=\eta_2=1$.
The unitarity of $S$ is ensured if $\cM$ satisfies
\begin{equation}
\cM_2^{-1} = K^{-1} - i P^2\,,
\label{eq:M2fromK}
\end{equation}
with $K$ a real, symmetric matrix.
Inserting the BB form of $S_2$ into \Cref{eq:M2fromS2} and comparing to \Cref{eq:M2fromK}, one finds
\begin{equation}
K^{-1} =  P \begin{pmatrix}
c_\epsilon^2 \cot\delta_1 + s_\epsilon^2 \cot\delta_2 & s_\epsilon c_\epsilon (\cot\delta_1 - \cot\delta_2)
\\
s_\epsilon c_\epsilon (\cot\delta_1 - \cot\delta_2) & c_\epsilon^2 \cot\delta_2 + s_\epsilon^2 \cot\delta_1
\end{pmatrix}  P
\,,
\label{eq:Kfromdelta}
\end{equation}
which has the expected decoupled limit when $\epsilon\to 0$.

To relate $K$ to the matrix in the quantization condition $\widehat \cK_{2,L}$, we follow the same steps
as above. The result \Cref{eq:M2Linv} remains valid, though now as a 2-d matrix equation.
Taking the $L\to\infty$ limit one now obtains
\begin{equation}
\cM_2^{-1} = \cK_2^{-1} - i H P^2\,,\quad
H = {\rm diag} \left( H_1(\bm k),\ H_2(\bm k) \right)\,,
\label{eq:M2fromKa}
\end{equation}
where $H_i$ are the appropriate cutoff functions for the two channels, as discussed above.
Comparing to \Cref{eq:M2fromK} one finds
\begin{equation}
\cK_2^{-1} = K^{-1} - i (1- H) P^2\,,
\label{eq:K2fromK}
\end{equation}
which is the generalization of \Cref{eq:K2ellinv}.

This discussion has assumed that both channels are above threshold. The analytic continuation below
threshold is done as shown in the definition of $\wt\rho$, \Cref{eq:trho}: $q^* \to i |q^*|$. 
This suffices to define the $P^2$ term in \Cref{eq:M2fromKa}.
As for $K^{-1}$, if one uses the expression \Cref{eq:Kfromdelta}, it would appear that we would need to
choose a branch of the square root in $P$. However, we know 
that the elements of $K$ are smooth functions of momenta, since it involves PV-regulated integrals.
Thus, in practice, it may be simplest to parametrize its three independent elements as analytic functions of 
the $q_i^2$.

\section{Form of matrix $\widehat{\cK}_{2,L}$ in charge basis}
\label{app:K2L}

Here we list the nonzero entries of $\widehat{\cK}_{2,L}$.
We use the shorthand notation
\begin{equation}
[K_{2,L}^{(ij\leftarrow mn;p)}]_{k'\ell' m', k \ell m} = 
\delta_{\bm k' \bm k} 2 \omega_{k_p} L^3 \delta_{\ell' \ell} \delta_{m' m} \cK_2^\ell( ij \leftarrow mn)\,,
\end{equation}
such that the right-hand side of \Cref{eq:K2Lmat36} is $K_{2,L}^{(\Kbar\pizero\leftarrow\Kminus\pizero;K^+)}$,
as well as the abbreviation 
\begin{equation}
K_{2,L}^{(ij;p)} \equiv K_{2,L}^{(ij\leftarrow ij;p)}
\end{equation}
for diagonal scattering.

The diagonal entries are 
\begin{multline}
\bigg\{ K_{2,L}^{(\Kplus\Kbar;\piminus)}, K_{2,L}^{(\Kplus\piminus;\Kbar)}, K_{2,L}^{(\Kbar\piminus;\Kplus)},
K_{2,L}^{(\Kplus\Kminus;\pizero)}, K_{2,L}^{(\Kplus\pizero;\Kminus)}, K_{2,L}^{(\Kminus\pizero;\Kplus)},
\\
K_{2,L}^{(\Kzero\Kbar;\pizero)}, K_{2,L}^{(\Kzero\pizero;\Kbar)}, K_{2,L}^{(\Kbar\pizero;\Kzero)},
K_{2,L}^{(\Kzero\Kminus;\piplus)}, K_{2,L}^{(\Kzero\piplus;\Kminus)}, K_{2,L}^{(\Kminus\piplus;\Kzero)},
\\
K_{2,L}^{([\piplus\piminus]_e;\eta)}, K_{2,L}^{([\piplus\piminus]_o;\eta)}, K_{2,L}^{(\piplus\eta;\piminus)},
K_{2,L}^{(\piminus\eta;\piplus)}, \tfrac12 K_{2,L}^{(\pizero\pizero;\eta)}, K_{2,L}^{(\pizero\eta;\pizero)}
\bigg\}
\end{multline}
We have augmented the notation with the subscripts $e$ and $o$ to indicated even and odd partial waves.

The nonzero offdiagonal entries lying above the diagonal are
\begin{align}
&\left[\widehat{\cK}_{2,L}\right]_{1, 15} = K_{2,L}^{(\Kplus \Kbar\leftarrow\piplus\eta; \piminus)},
&\left[\widehat{\cK}_{2,L}\right]_{2, 8} = K_{2,L}^{(\Kplus \piminus\leftarrow\Kzero\pizero; \Kbar)},
\\
&\left[\widehat{\cK}_{2,L}\right]_{3, 6} = K_{2,L}^{(\Kbar\piminus\leftarrow\Kminus\pizero;\Kplus)},
&\left[\widehat{\cK}_{2,L}\right]_{4, 7} = K_{2,L}^{(\Kplus \Kminus\leftarrow\Kzero\Kbar; \pizero)},
\\
&\left[\widehat{\cK}_{2,L}\right]_{4, 18} = K_{2,L}^{(\Kplus \Kminus\leftarrow\pizero\eta; \pizero)},
&\left[\widehat{\cK}_{2,L}\right]_{5, 11} = K_{2,L}^{(\Kplus\pizero\leftarrow\Kzero\piplus;\Kminus)},
\\
&\left[\widehat{\cK}_{2,L}\right]_{7, 18} = K_{2,L}^{(\Kzero \Kbar\leftarrow\pizero\eta; \pizero)},
&\left[\widehat{\cK}_{2,L}\right]_{9, 12} = K_{2,L}^{(\Kbar \pizero\leftarrow\Kminus\piplus ; \Kzero)},
\\
&\left[\widehat{\cK}_{2,L}\right]_{10, 16} = K_{2,L}^{(\Kzero\Kminus\leftarrow\piminus\eta; \piplus)},
&\left[\widehat{\cK}_{2,L}\right]_{13, 17} = \sqrt{\tfrac12} K_{2,L}^{([\piplus\piminus]_e\leftarrow\pizero\pizero; \eta)},
\end{align}
while those below the diagonal are determined by the result that the matrix is symmetric.
Symmetry follows from PT symmetry, which implies that the K matrices for $ij \leftarrow mn$ and
$mn \leftarrow ij$ are equal.

The factors of $1/2$ and $\sqrt{1/2}$ in the expressions above are symmetry factors that
follow from the considerations of BS2.

\section{Isospin decompositions of two-particle K matrices}
\label{app:K2isospin}

In the following, the superscript $\ell$ is dropped, and we make use of the form of the triplet and
singlet $K \overline K$ pairs,
\begin{align}
[K \bar K]_1 &= \left\{ - \Kplus \Kbar,
\sqrt{\tfrac12} ( \Kplus \Kminus - \Kzero\Kbar ),
\Kzero\Kminus \right\}\,,
\\
[K \bar K]_0 &= \sqrt{\tfrac12} ( \Kplus \Kminus + \Kzero\Kbar )\,.
\end{align}

For $\pi\pi$ scattering we have
\begin{align}
\cK_2([\pi^+\pi^-]_e \leftarrow [\pi^+\pi^-]_e) &= \frac16 \left(\cK_2^{\pi\pi,I=2} + 2 \cK_2^{\pi\pi,I=0} \right)\,.
\\
\cK_2(\pi^0\pi^0 \leftarrow \pi^0\pi^0) &= \frac13 \left(2\cK_2^{\pi\pi,I=2} + \cK_2^{\pi\pi,I=0} \right)\,,
\\
\cK_2(\pi^0\pi^0 \leftrightarrow [\pi^+\pi^-]_e) &= \frac13 \left(\cK_2^{\pi\pi,I=2} - \cK_2^{\pi\pi,I=0} \right)\,,
\\
\cK_2([\pi^+\pi^-]_o \leftrightarrow [\pi^+\pi^-]_o) &= \frac12 \cK_2^{\pi\pi,I=1}\,.
\label{eq:pippimo}
\end{align}
To obtain these results, care must be taken with the meaning of the subscripts $e$ and $o$.
For example, to obtain the last line, we use the fact that the $I=1,I_3=0$ state is
$(\pi^+\pi^0-\pi^-\pi^+)/\sqrt2$, and determine the contribution of all contractions.
Even partial waves cancel, while odd waves come with a factor of $(1/\sqrt2)^2 \times 4=2$.
Since $[\pi^+\pi^-]_o \leftarrow [\pi^+\pi^-]_o$ means ``take the odd partial waves of the
amplitude for $\pi^+\pi^-\leftarrow \pi^+\pi^-$", this leads to the factor of $1/2$ on the right-hand side
of \Cref{eq:pippimo}.

For $\pi \eta$ scattering we have
\begin{equation}
\cK_2(\pi^+\eta \leftarrow \pi^+\eta) = \cK_2(\pi^0\eta \leftarrow \pi^0\eta) 
= \cK_2(\pi^-\eta \leftarrow \pi^-\eta) = \cK_2^{\pi\eta,I=1}\,,
\end{equation}
while for $\pi \eta \leftrightarrow K \overline K$ we find
\begin{multline}
-\cK_{2,L}(\pi^+\eta \leftrightarrow K^+ \Kbar)
=
\cK_{2,L}(\pi^-\eta \leftrightarrow K^0 K^-)
=
\\
=
- \sqrt2 \cK_{2,L}(\pi^0\eta \leftrightarrow K^0 \Kbar)
=
\sqrt2 \cK_{2,L}(\pi^0\eta \leftrightarrow K^+ K^-)
=
\cK_{2,L}^{\pi\eta \leftrightarrow K\bar K, I=1}
\,.
\end{multline}
For $K\pi$ scattering, the results are
\begin{align}
\cK_2(K^+\pi^- \leftarrow K^+\pi^- ) &= \cK_2(K^0\pi^+  \leftarrow K^0\pi^+ )\
= \frac13 \left(\cK_2^{K\pi,I=3/2} +2 \cK_2^{K\pi,I=1/2} \right)\,,
\\
\cK_2(K^0 \pi^0\leftarrow K^0\pi^0) &= \cK_2(K^+\pi^0  \leftarrow K^+\pi^0 )
= \frac13 \left(2\cK_2^{K\pi,I=3/2} + \cK_2^{K\pi,I=1/2} \right)
\,,
\\
\cK_2(K^+ \pi^-\leftrightarrow K^0\pi^0) &= \cK_2( K^0\pi^+ \leftrightarrow K^+ \pi^0)
= \frac{\sqrt2}3 \left(\cK_2^{K\pi,I=3/2} - \cK_2^{K\pi,I=1/2} \right)
\,.
\end{align}
Those for $\overline K \pi$ scattering can then be obtained using charge conjugation,
keeping in mind that, for standard conventions, $C \pi^+=-\pi^-$, $C K^+=K^-$, and
$C K^0 = \Kbar$. We find
\begin{align}
\cK_2(K^-\pi^+ \leftarrow K^-\pi^+) &=\cK_2(\Kbar\pi^- \leftarrow  \Kbar\pi^-)
= \frac13 \left(\cK_2^{K\pi,I=3/2} +2 \cK_2^{K\pi,I=1/2} \right)\,,
\\
\cK_2(\Kbar\pi^0 \leftarrow \Kbar\pi^0) &=\cK_2(K^-\pi^0 \leftarrow K^-\pi^0)
= \frac13 \left(2\cK_2^{K\pi,I=3/2} + \cK_2^{K\pi,I=1/2} \right)\,,
\\
\cK_2(K^-\pi^+ \leftrightarrow \Kbar\pi^0) &=\cK_2(\Kbar \pi^-  \leftrightarrow K^- \pi^0)
= \frac{\sqrt2}3 \left(- \cK_2^{K\pi,I=3/2} + \cK_2^{K\pi,I=1/2} \right)\,.
\end{align}
Finally, for $K \overline K$ scattering, we have
\begin{align}
\cK_2(K^+\Kbar \leftarrow K^+\Kbar) &= \cK_2(K^0 K^- \leftarrow K^0 K^-) 
= \cK_2^{K\bar K,I=1}  \,,
\\
\cK_2(K^+ K^- \leftarrow K^+ K^-) 
&= \cK_2(K^0 \Kbar \leftarrow K^0 \Kbar) 
= \frac12\left(\cK_2^{K\bar K,I=0} + \cK_2^{K\bar K,I=1} \right) 
\,,
\\
\cK_2(K^+ K^- \leftrightarrow K^0 \Kbar) 
&= \frac12\left(\cK_2^{K\bar K,I=0} - \cK_2^{K\bar K,I=1} \right)\,.
\end{align}

\section{Conversion from charge to isospin basis for 18-d matrices}
\label{app:chtoiso}

We can convert the 18-d matrices to the isospin basis \Cref{eq:Ichannels18} by
means of conjugating by the $18\times 18$ orthogonal matrix $C_{{\rm ch}\to{\rm iso}}^{(18)}$:
\begin{equation}
M_{\rm iso}^{(18)} = C_{{\rm ch}\to{\rm iso}}^{(18)} M_{\rm ch}^{(18)} \left[C_{{\rm ch}\to{\rm iso}}^{(18) }\right]^{-1}\,.
\label{eq:chtoiso}
\end{equation}
To obtain this matrix we simply expand the entries of  \Cref{eq:Ichannels18} using isospin Clebsch-Gordon
coefficients, and then express in terms of the charge basis, \Cref{eq:pairspect0}.
The only subtlety is the treatment of $[\piplus\piminus]_{e/o}$.
In order for $C_{{\rm ch}\to{\rm iso}}^{(18)}$ to correspond to a change of basis, it must be unitary,
and this requires maintaining normalizations.
Thus we must consider $[\piplus\piminus]_{e/o}$ as corresponding to
$[\pi^+ \pi^- \pm \pi^- \pi^+]/\sqrt2$.

The conversion matrix is too large to present in normal format, so we present it row by row in a compact notation.
We begin with the first five rows, which are those corresponding to $I=2$, and are given by
\begin{align}
[[K\bar K]_1 \pi]_2:&\ \sqrt{\tfrac16} \left(-1,0,0,\sqrt2,0,0,-\sqrt2,0,0,1,\ket0_6 \right)\,,
\\
[[K \pi]_{3/2} \bar K]_2:&\ \sqrt{\tfrac16} \left(0,-1,0,0,\sqrt2,0,0,-\sqrt2,0,0,1,0 ,\ket0_{6}\right)\,,
\\
[[\bar K\pi]_{3/2}  K]_2:&\ \sqrt{\tfrac16} \left(0,0,-1,0,0,\sqrt2,0,0,-\sqrt2,0,0,1,\ket0_{6}\right)\,,
\\
[[\pi \pi]_2\eta]_2:&\ \sqrt{\tfrac13} \left(\ket{0}_{12},1,0,0,0,\sqrt2, 0\right)\,
\\
[[\pi \eta]_1 \pi]_2:&\  \sqrt{\tfrac16} \left( \ket{0}_{12}, 0,0,1,1, 0, 2 \right)\,,
\end{align}
where $\ket0_n$ is the vector consisting of $n$ zeros.

The middle eight rows, corresponding to $I=1$, are
\begin{align}
[[K\bar K]_1 \pi]_1:&\ \sqrt{\tfrac12} \left(-1,0,0,0,0,0,0,0,-1,0,0,\ket0_{6} \right)\,,
\\
[[K\bar K]_0 \pi]_1:&\ \sqrt{\tfrac12} \left(0,0,0,1,0,0,1,0,0,0,0,0,\ket0_{6} \right)\,,
\\
[[K \pi]_{3/2} \bar K]_1:&\  \sqrt{\tfrac16} \left(0,1,0,0,\sqrt2,0,0,\sqrt2,0,0,1,0,\ket0_{6} \right)\,,
\\
[[K \pi]_{1/2} \bar K]_1:&\  \sqrt{\tfrac16} \left(0,-\sqrt2,0,0,1,0,0,1,0,0,-\sqrt2,0,\ket0_{6}\right)\,,
\\
[[\bar K\pi]_{3/2}  K]_1:&\ \sqrt{\tfrac16} \left(0,0,1,0,0,-\sqrt2,0,0,-\sqrt2,0,0,1,\ket0_{6} \right)\,,
\\
[[\bar K\pi]_{1/2}  K]_1:&\ \sqrt{\tfrac16} \left(0,0,-\sqrt2,0,0,-1,0,0,-1,0,0,-\sqrt2,\ket0_{6} \right)\,,
\\
[[\pi\pi]_1\eta]_1:&\ \left(\ket{0}_{12}, 0,1,0,0,0, 0\right)\,,
\label{eq:I1ai}
\\
[[\pi \eta]_1 \pi]_1:&\  \sqrt{\tfrac12} \left(\ket{0}_{12}, 0,0,1,-1, 0, 0 \right)\,.
\end{align}

The last five rows, corresponding to $I=0$, are
\begin{align}
[[K\bar K]_1 \pi]_0:&\ \sqrt{\tfrac16} \left(-\sqrt2,0,0,-1,0,0,1,0,0,\sqrt2,0,0,\ket0_{6}\right)\,,
\\
[[K \pi]_{1/2} \bar K]_0:&\  \sqrt{\tfrac16} \left(0,\sqrt2,0,0,1,0,0,-1,0,0,-\sqrt2,0,\ket0_{6} \right)\,,
\\
[[\bar K\pi]_{1/2}  K]_0:&\ \sqrt{\tfrac16} \left(0,0,\sqrt2,0,0,1,0,0,-1,0,0,-\sqrt2,\ket0_{6} \right)\,,
\\
[[\pi\pi]_0\eta]_0:&\ \sqrt{\tfrac13} \left(\ket{0}_{12},\sqrt2,0,0,0,-1, 0 \right)\,
\\
[[\pi \eta]_1 \pi]_0:&\  \sqrt{\tfrac13} \left(\ket{0}_{12},0,0,1,1, 0, -1\right)\,,
\end{align}

\section{Quadratic terms in the threshold expansion of $\Kdf$}
\label{app:quadratic}

Here we collect the quadratic terms in the threshold expansion developed in \Cref{sec:Kdfparam},
and sketch the group-theoretic analysis that allows one to determine the number of terms with
given symmetry properties.
Rather than give names to the coefficients we simply list the independent terms, each
of which will be multiplied by an independent real coefficient.

For the $I=0,2$ diagonal K matrices we find 11 terms, one choice of basis for which is,
\begin{equation}
\begin{split}
\mathcal{K}^{[I],xx}_\text{df,3}:& \
 \Delta^2,\ (\Delta_{1} \Delta_{2} + {\Delta'}_{1} {\Delta'}_{2}),\
(\tilde {t}_{11} \tilde{t}_{22} + \tilde{t}_{12} \tilde {t}_{21}),\
(\Delta_{1} + \Delta_{2}) ({\Delta'}_{1} + {\Delta'}_{2}),\
\\ &\
(\tilde{t}_{11} + \tilde{t}_{22}) (\tilde{t}_{12} + \tilde{t}_{21}),\
(\Delta_{1}^2 + {\Delta'}_{1}^2 + \Delta_{2}^2 + {\Delta'}_{2}^2),\
\\ &\
\Delta (\Delta_{1} + {\Delta'}_{1} + \Delta_{2} + {\Delta'}_{2}),\
(\tilde{t}_{11}^2 + \tilde{t}_{12}^2 + \tilde{t}_{21}^2 + \tilde{t}_{22}^2), \
\Delta (\tilde {t}_{11} + \tilde{t}_{12} + \tilde {t}_{21} + \tilde{t}_{22}),\
\\ &\
(\tilde {t}_{11} (\Delta'_{2} + {\Delta}_{2}) + \tilde {t}_{12} ({\Delta'}_{2} + \Delta_{1}) 
+ \tilde{t}_{21} (\Delta'_{1} + {\Delta}_{2}) + \tilde{t}_{22} (\Delta'_{1} + {\Delta}_{1}))
\\ &\
(\tilde{t}_{11} (\Delta'_{1} + {\Delta}_{1}) + \tilde{t}_{12} (\Delta'_{1} + {\Delta}_{2}) + 
\tilde{t}_{21} ({\Delta'}_{2} + \Delta_{1}) + \tilde{t}_{22} (\Delta'_{2} + {\Delta}_{2}))\,, 
\end{split}
\label{eq:KdfthrI02xxquad}
\end{equation}
while for the offdiagonal case we find 17 terms,
\begin{equation}
\begin{split}
\mathcal{K}^{[I],ab}_\text{df,3}: &\
\Delta^2,\
\Delta_{1} \Delta_{2},\
{\Delta'}_{1} {\Delta'}_{2},\
(\Delta_{1}^2 + \Delta_ {2}^2),\
({\Delta'}_{1}^2 + {\Delta'}_{2}^2),\
\Delta (\Delta_{1} + \Delta_{2}),\
 \\ &\
 \Delta ({\Delta'}_{1} + {\Delta'}_{2}),\
 (\tilde{t}_{11} \tilde{t}_{12} + \tilde{t}_{21} \tilde{t}_{22}),\
 (\tilde{t}_{11} \tilde{t}_{21} + \tilde{t}_{12} \tilde{t}_{22}),\
 (\tilde{t}_{11} \tilde{t}_{22} + \tilde{t}_{12} \tilde{t}_{21}),\
\\ & \
(\Delta_{1} + \Delta_{2}) ({\Delta'}_{1} + {\Delta'}_{2}),\
(\tilde{t}_{11}^2 + \tilde{t}_{12}^2 + \tilde{t}_{21}^2 + \tilde{t}_{22}^2),\
\Delta (\tilde{t}_{11} + \tilde{t}_{12} + \tilde{t}_{21} + \tilde{t}_{22}),\
\\ & \
((\tilde{t}_{11} + \tilde{t}_{21}) \Delta_{1}  + (\tilde{t}_{12} + \tilde{t}_{22}) \Delta_{2}),\
((\tilde{t}_{11} + \tilde{t}_{21}) \Delta_{2} + (\tilde{t}_{12} + \tilde{t}_{22}) \Delta_{1} ),\
\\ & \
({\Delta'}_{1} (\tilde{t}_{11} + \tilde{t}_{12}) + {\Delta'}_{2} (\tilde{t}_{21} + \tilde{t}_{22})),\
({\Delta'}_{2} (\tilde{t}_{11} + \tilde{t}_{12}) + {\Delta'}_{1} (\tilde{t}_{21} + \tilde{t}_{22}))\,.
\end{split}
\label{eq:KdfthrI02abquad}
\end{equation}

For $I=1$, the $aa$ quadratic terms are as in \Cref{eq:KdfthrI02xxquad}, while those for
$bb$ and $cc$ involve the following 6 terms
\begin{equation}
\begin{split}
\mathcal{K}^{[I=1],yy}_\text{df,3}: &\
(\tilde{t}_{12} \tilde{t}_{21} - \tilde{t}_{11} \tilde{t}_{22}),\
(\Delta_{1} - \Delta_{2}) ({\Delta'}_{1} - {\Delta'}_{2}),
\\ & \
(\tilde{t}_{11}^2 - \tilde{t}_{12}^2 - \tilde{t}_{21}^2 + \tilde{t}_{22}^2),\
\Delta (\tilde{t}_{11} - \tilde{t}_{12} - \tilde{t}_{21} + \tilde{t}_{22}),
\\ & \
( \Delta'_{1} (\tilde{t}_{11} - \tilde{t}_{12})  +  \Delta'_{2} (\tilde{t}_{22} - \tilde{t}_{21}) 
+ (\tilde{t}_{11} - \tilde{t}_{21}) \Delta_{1}+ (\tilde{t}_{22} - \tilde{t}_{12})\Delta_{2}  ),
\\ & \
( \Delta'_{1} (\tilde{t}_{22} - \tilde{t}_{21})  +  \Delta'_{2} (\tilde{t}_{11} - \tilde{t}_{12}) 
+ (\tilde{t}_{22} - \tilde{t}_{12}) \Delta_{1}+ (\tilde{t}_{11} - \tilde{t}_{21})\Delta_{2}  ).
\end{split}
\label{eq:KdfthrI1yyquad}
\end{equation}
As for the offdiagonal $I=1$ contributions, we find 10 quadratic terms for $ay$, with $y=b$ or $c$
\begin{equation}
\begin{split}
\mathcal{K}^{[I=1],ay}_\text{df,3}(\{k'\},\{k\}): &\
(\tilde {t}_{11} \tilde{t}_{21} - \tilde {t}_{12} \tilde {t}_{22}),\
(\Delta'_ {1} + \Delta'_ {2}) ({\Delta}_ {1} - {\Delta}_ {2}),\
 ({\Delta'}_ {1}^2 + {\Delta'}_{2}^2),\ 
 ({\Delta}_ {1}^2 - {\Delta}_ {2}^2),\
\\ & \
 (\tilde {t}_{11}^2 - \tilde {t}_{12}^2 + \tilde{t}_{21}^2 - \tilde {t}_{22}^2),\
\Delta (\tilde {t}_{11} - \tilde{t}_{12} + \tilde {t}_{21} - \tilde{t}_{22}),\
\\ & \
((\tilde{t}_{12} + \tilde{t}_{22}) \Delta_1 - (\tilde{t}_{11} + \tilde {t}_{21}) \Delta_2),\
((\tilde{t}_{11} + \tilde {t}_{21}) \Delta_1 - (\tilde{t}_{12} + \tilde {t}_{22}) \Delta_1),\
\\ & \ 
(\Delta'_{1} (\tilde {t}_{11} - \tilde {t}_{12}) + \Delta'_{2} (\tilde{t}_{21} - \tilde{t}_{22})),\
(\Delta'_{2} (\tilde {t}_{11} - \tilde {t}_{12}) + \Delta_{1} (\tilde{t}_{21} - \tilde{t}_{22}))\,,
\end{split}
\label{eq:KdfthrI1ayquad}
\end{equation}
and 8 terms for $bc$
\begin{equation}
\begin{split}
\mathcal{K}^{[I=1],bc}_\text{df,3}: &\
(\tilde{t}_{12} \tilde{t}_{21} - \tilde{t}_{11} \tilde{t}_{22}),\
(\Delta_{1} - \Delta_{2}) ({\Delta'}_{1} - {\Delta'}_{2}),\
\\ & \
(\tilde{t}_{11}^2 - \tilde{t}_{12}^2 - \tilde{t}_{21}^2 + \tilde{t}_{22}^2),\
\Delta (\tilde{t}_{11} - \tilde{t}_{12} - \tilde{t}_{21} + \tilde{t}_{22}),
\\ & \
( \Delta'_{1} (\tilde{t}_{11} - \tilde{t}_{12})  +  \Delta'_{2} (\tilde{t}_{22} - \tilde{t}_{21})),\ 
((\tilde{t}_{11} - \tilde{t}_{21}) \Delta_{1}+ (\tilde{t}_{22} - \tilde{t}_{12})\Delta_{2}  ),
\\ & \
( \Delta'_{1} (\tilde{t}_{22} - \tilde{t}_{21})  +  \Delta'_{2} (\tilde{t}_{11} - \tilde{t}_{12}) ),\
((\tilde{t}_{22} - \tilde{t}_{12}) \Delta_{1}+ (\tilde{t}_{11} - \tilde{t}_{21})\Delta_{2}  ).
\end{split}
\label{eq:KdfthrI1bcquad}
\end{equation}

The counting of terms can be done using a group theoretic method introduced in 
appendix B of Ref.~\cite{\kdfnloall}.
This has the advantage over direct enumeration of avoiding the possibility of missing terms.
The method is based on the observation that all terms can be built from products of the $\wt t_{ij}$,
as can be seen from the relations \Cref{eq:constraints}.
These 9 objects transform under the symmetry group $S_2 \times S'_2$, where $S_2$ is the
permutation group of two elements, here generated by the interchange initial momenta $k_1 \leftrightarrow k_2$,
while $S'_2$ is the corresponding group for final state interchange.
This is the symmetry group that is applicable for the off-diagonal elements of $\Kdf$.
For the diagonal elements, PT symmetry enlarges the group to $(S_2 \times S'_2)\rtimes Z_2$,
with the $Z_2$ interchanging initial and final states. The $Z_2$ acts nontrivially---interchanging the two
$S_2$s---so the combination involves a semidirect product.

The method is now to decompose the 9 objects $\wt t_{ij}$ and the 45 objects $\wt t_{ij} \wt t_{k\ell}$ into
irreducible representations (irreps) of the appropriate group. We begin with the simpler $S_2 \times S'_2$ case.
Here there are four 1-d irreps, labeled $(1,1)$, $(1,-1)$, $(-1,1)$ and $(-1,-1)$ according to symmetry/antisymmetry
under the two subgroups. Determining the characters of $\wt t_{ij}$, we find that these 9 objects decompose as
\begin{equation}
4 \times (1,1) \oplus 2 \times (1, -1) \oplus 2\times (-1,1) \oplus (-1,-1)\,.
\end{equation}
The 4 singlets lead to the four linear terms in $\cK_{\rm df,3}^{[I=0,2]ab}$, \Cref{eq:KdfthrI02ab}, 
the two $(-1,1)$ irreps lead to the two linear terms in $\cK_{\rm df,3}^{[1]ab}$, \Cref{eq:KdfthrI1ay},
and the single $(-1,-1)$ irrep leads to the single linear term in $\cK_{\rm df,3}^{[1]bc}$, \Cref{eq:KdfthrI1bc}.
A similar exercise for the $\wt t_{ij} \wt t_{k\ell}$ finds
\begin{equation}
17 \times (1,1) \oplus 10 \times (1, -1) \oplus 10\times (-1,1) \oplus 8(-1,-1)\,.
\end{equation}
This determines the number of independent quadratic terms, and is consistent with the results quoted earlier
in this section.

Now we turn to the diagonal cases, which have the larger $(S_2 \times S'_2)\rtimes Z_2$ symmetry.
The group can be represented by $4-d$ matrices that act on the momentum vector
$\{k_1, k_2, k'_1, k'_2\}$, and is generated by the block matrices
\begin{equation}
s_2 =\begin{pmatrix} \sigma_1 & 0 \\ 0 & 1 \end{pmatrix},\ \
s'_2 =\begin{pmatrix} 1 & 0 \\ 0 & \sigma_1 \end{pmatrix},\ \ {\rm and} \ \
z_2 = \begin{pmatrix} 0 & 1 \\ 1 & 0 \end{pmatrix}\,,
\end{equation}
with $\sigma_1$ the usual Pauli matrix.
There are 5 conjugacy classes, with elements $\{1\}$, $\{s_2, s'_2\}$,
$\{s_2 s'_2\}$, $\{z_2, s_2 z_2 s_2\}$, and $\{z_2 s_2, z_2 s_2'\}$.
There are correspondingly 5 irreps, four with $d=1$ and one with $d=2$.
The character table is shown in \Cref{tab:char}, with rather arbitrary names chosen for the irreps.

The character vector of the $\wt t_{ij}$ is
$\chi_t = \{9,3,1,3,1\}$, from which one finds the decomposition
\begin{equation}
3 \times {\bf 1} \oplus {\bf 1a} \oplus {\bf 1b} \oplus 2 \times {\bf 2}\,.
\end{equation}
The three singlets correspond to the three linear terms in $\cK_{\rm df,3}^{[I=0,2],xx}$, \Cref{eq:KdfthrI02xx},
while the ${\bf 1a}$ irrep is that which is fully antisymmetric, and thus leads to the single term in
$\cK_{\rm df,3}^{[I=1],yy}$, \Cref{eq:KdfthrI1yy}.

For $\wt t_{ij} \wt t_{k\ell}$, the character vector is $\chi_{t^2} = \{45,9,5,9,1\}$, leading to the decomposition
\begin{equation}
11 \times {\bf 1} \oplus 6\times {\bf 1a} \oplus 6\times {\bf 1b} \oplus 2\times {\bf 1c} \oplus 10 \times {\bf 2}\,.
\end{equation}
This leads to the 11 fully symmetric quadratic terms in \Cref{eq:KdfthrI02xxquad} above,
and the 6 fully antisymmetric terms in \Cref{eq:KdfthrI1yyquad}

\begin{table}[htp]
\caption{Character table of $(S_2\times S'_2)\rtimes Z_2$.}
\begin{center}
\begin{tabular}{c||c|c|c|c|c}
Class & $1$ & $s_2$ & $s_2 s'_2$ & $z_2$ & $z_2 s_2$\\  
\hline
Dim & 1 & 2 & 1 & 2 & 2\\
\hline
{\bf 1}   & 1 & 1 & 1 & 1 & 1  \\
{\bf 1a} & 1 & -1 & 1 & 1 & -1 \\
{\bf 1b} & 1 & 1 & 1 & -1 & -1 \\
{\bf 1c} & 1 & -1 & 1 & -1 & 1 \\
{\bf 2} & 2 & 0 & -2 & 0 &  0 
\end{tabular}
\end{center}
\label{tab:char}
\end{table}

\section{Symmetrization of the quantization condition}
\label{app:sym}

In this appendix we sketch how the asymmetric form of the three-particle quantization condition,
\Cref{eq:QC3asym}, can be manipulated into the symmetric form, \Cref{eq:QC3sym}.
This is done using symmetrization identities, which are valid only up to
exponentially-suppressed corrections. This is sufficient for our purposes, however, since the
derivation of all forms of the quantization condition discards such corrections.

For simplicity, we consider a single three-particle channel with distinguishable particles, allowing us to
use several results from ref.~\cite{\BSnondegen} (BS3).
The generalization to identical particles, $2+1$ systems, or multiple three-particle
channels, is straightforward, since the symmetrization identities take the same form in all cases.
For brevity, we also drop carets in this appendix.

The asymmetric form of the quantization condition can be rewritten as
\begin{equation}
\det M = 0\,,\quad M = 1 + F_G(\cK_{2,L}+ \cK_{\rm df,3}^{(u,u)})\,.
\end{equation}
Noting that $\cD_{23,L}^{(u,u)}$, \Cref{eq:D23Luu}, can be written as
\begin{equation}
\cD_{23,L}^{(u,u)} = \frac1{\cK_{2,L}^{-1} + F_G}\,,
\end{equation}
we find
\begin{equation}
M = \frac1{1- F_G \cD_{23,L}^{(u,u)}}  M'\,,\quad
M'=\left[1 + (1-F_G \cD_{23,L}^{(u,u)}) F_G \cK_{\rm df,3}^{(u,u)} \right]\,.
\end{equation}
Since energy levels must depend on the three-particle K matrix, assuming that it has a general form,
we can rewrite the quantization condition as $\det M'=0$. We note that $M'$ has the same form as the
denominator in eq.~(79) of BS3.

The next step is to use the algebraic result given in eq.~(D9) of BS3,
\begin{align}
\cK_{\rm df,3}^{(u,u)} M'^{-1} &= (1 + \cK_{2,L} \overrightarrow{I_G} ) \cT_L (1 + \overleftarrow{I_G} \cK_{2,L})\,,
\label{eq:G4}
\\
\cT_L &= \cK' \frac1{1 + \left[F_G-I_{FG} - (F_G - \overleftarrow{I_G}) \cD_{23,L}^{(u,u)}
(F_G - \overrightarrow{I_G}) \right] \cK'}\,,
\\
\cK' &= Z \frac1{1+ (I_{FG} + \overleftarrow{I_G} \cK_{2,L} \overrightarrow{I_G} ) Z}\,,
\label{eq:G6}
\\
Z &= \frac1{1+\cK_{2,L} \overrightarrow{I_G} } \cK_{\rm df,3}^{(u,u)}\frac1{1+\overleftarrow{I_G} \cK_{2,L}}\,.
\label{eq:G7}
\end{align}
where $I_{FG}$, $\overrightarrow{I_G}$ and $\overleftarrow{I_G}$ are integral operators that enter
the symmetrization identities given in eqs.~(100-102) of BS3, which, in shorthand form, are
\begin{equation}
F_G - I_{FG} = \frac13 \overleftarrow{\cS} F \overrightarrow{\cS}\,,\quad
F_G-\overrightarrow{I_G} = F \overrightarrow{\cS}\,,\quad
F_G-\overleftarrow{I_G} =  \overleftarrow{\cS} F\,.
\end{equation}
Here $\overrightarrow{\cS}$ and $\overleftarrow{\cS}$ are symmetrization operators,
defined in eq.~(98) of BS3.

Using these identities we find
\begin{align}
\cT_L &= \cK' - \cK' \overleftarrow{\cS} F_3 \frac1{1+\Kdf F_3} \overrightarrow{\cS} \cK'\,,
\label{eq:G9}
\\
F_3 &= \frac{F}3 - F \cD_{23,L}^{(u,u)} F = 
\frac{F}3 - F \frac1{\cK_{2,L}^{-1} + F_G} F\,,
\\
\Kdf &= \overrightarrow{\cS} \cK' \overleftarrow{\cS}\,.
\end{align}
This leads to the desired symmetric form of the quantization condition in the following manner.
As noted above, the quantization condition can be written $\det M'=0$,
which implies that the left-hand side of \Cref{eq:G4} must satisfy $\det(\cK_{\rm df,3}^{(u,u)} M'^{-1}) = \infty$.
Now, in the right-hand side, the integral operators sew the external factors
of $\cK_{2,L}$ to the central $\cT_L$ and do not lead to divergences. These can only arise from
$\cT_L$ itself, so that the quantization condition can be rewritten as $\det \cT_L=\infty$.
Using \Cref{eq:G9}, we then observe that this divergence can arise only in the second term on the
right-hand side, leading to the standard symmetric form of the quantization condition,
\begin{equation}
\det(1 + \Kdf F_3) = 0\,.
\end{equation}
The symmetrization operators on both sides of $\cK'$ ensure that $\Kdf$ is symmetric.
Tracing back through \Cref{eq:G6,eq:G7} we see that $\cK'$ is obtained from the K matrix
that enters the asymmetric form of the quantization condition, $\cK_{\rm df,3}^{(u,u)}$,
by an infinite series involving attaching factors of $\cK_{2,L}$ with the integral operators.
An important point is that these attachments convert one infinite-volume quantity into another.

\bibliographystyle{JHEP}
\bibliography{ref.bib}

\end{document}